\documentclass[11pt]{article}
\usepackage[english]{babel}\usepackage{setspace}
\usepackage[letterpaper,top=2cm,bottom=2cm,left=3cm,right=3cm,marginparwidth=1.75cm]{geometry}\usepackage[title]{appendix}
\usepackage{amsthm,dirtytalk,natbib,bm,enumerate} \usepackage{float} \newtheorem{proposition}{Proposition} \newtheorem{lemma}{Lemma}\newtheorem{corollary}{Corollary} \newtheorem{theorem}{Theorem} \newtheorem{assumption}{Assumption} \theoremstyle{definition} \newtheorem{remark}{Remark} \newtheorem{definition}{Definition} \usepackage{graphicx,amsfonts,mathtools,soul,comment}
\usepackage{hyperref,xcolor,booktabs,threeparttable,amsmath}
\definecolor{ahjcolor}{rgb}{0.0, 0.13, 0.40}	
\hypersetup{colorlinks	=true,									 linkcolor	=ahjcolor,									         urlcolor	=ahjcolor,									            citecolor	=ahjcolor,            bookmarksnumbered=true}
\DeclareMathOperator*{\argmin}{arg\,min}
\title{Locally Robust Inference with Outcomes Predicted by Machine Learning\thanks{This paper is based on the third chapter of Puerta-Cuartas's doctoral thesis at Universidad Carlos III de Madrid, defended on June 30, 2026. Escanciano acknowledges support from grants PID2021-127794NB-I00 and CEX2021-001181-M, funded by MCIN/AEI/10.13039/501100011033, PID2025-170712NB-I00, funded by  ICIU/AEI/10.13039/501100011033, and Comunidad de Madrid, grants EPUC3M11 (VPRICIT) and H2019/HUM-589. The views expressed in this paper are those of the authors and do not necessarily coincide with those of the Bank of Spain or the Eurosystem.}}
\author{Juan Carlos Escanciano \\
\textit{Universidad Carlos III de Madrid} \and Alejandro Puerta-Cuartas\thanks{Corresponding author: \href{E-mail: alejandro.puerta@bde.es}{E-mail: alejandro.puerta@bde.es}.}\\
\textit{Banco de España} }
\begin{document} \onehalfspacing
\maketitle 
\begin{abstract} Many parameters of interest in economics depend on outcomes generated by machine learning (ML). These generated outcomes are subsequently used in standard econometric estimation and typically treated as observed, a practice that threatens the validity of downstream inference. Relevant examples include the partially linear model with an ML-predicted outcome, productivity persistence, and inequality of opportunity with life-cycle-corrected income. This paper develops a three-step locally robust GMM framework for valid inference when the dependent variable is generated by ML. We construct a debiased moment condition that augments the identifying moment with first- and second-step influence functions, making inference locally insensitive to regularization and model selection biases at both steps. We establish consistency and asymptotic normality of the locally robust estimator and illustrate its finite-sample performance through simulations. We then estimate inequality of opportunity in Spain, where incomes observed at different ages are subject to life-cycle bias. Correcting income for this bias with type-specific age profiles estimated by ML lowers inequality of opportunity over the working life by 10\% to 14\% relative to raw income, whereas treating the corrected incomes as observed overstates it by up to 15\%.

\vspace{1em}
 \end{abstract}
\newpage \onehalfspacing
\section{Introduction}
The rapid growth in rich, high-dimensional data availability has created new opportunities for empirical economic research. Examples include unstructured text, images, and satellite imagery, from which machine learning (ML) methods have become increasingly common. In particular, ML is often used to construct key variables, whether regressors or outcomes, that are subsequently employed using standard econometric techniques. However, these constructed variables, known as generated variables, depend fundamentally on first-stage estimation procedures, threatening valid inference.\par 
Generated dependent variables arise in many empirical applications, where the parameter of interest depends not on a directly observed outcome but on one constructed in a preliminary estimation step. A growing part of the empirical literature already relies on ML to construct outcomes of interest. For instance, researchers use remote sensing and other high-dimensional data to construct variables such as consumption, GDP, and agricultural yields \citep{jean2016combining, henderson2012measuring, burke2017satellite, graetz2018mapping}. In such contexts, the parameter of interest is typically the effect of some key variable on the constructed outcome, controlling for confounders. In other settings, practitioners rely on parametric specifications to construct the outcome. For inequality of opportunity measurement, e.g., \cite{brunori2025inequality}, income is purged of life-cycle variation or cohort fixed effects before estimating how much of its remaining variation is attributable to circumstances beyond individual control, such as parental background or place of birth. In partial insurance models, e.g., \cite{blundell2008consumption}, consumption is predicted from auxiliary data before being used to study how households smooth income shocks. Whether the outcome is constructed by machine learning or by parametric methods, its preliminary estimation threatens the validity of inference. \par 
Machine learning offers an appealing choice for constructing generated outcomes. Specifically, it handles high-dimensional covariates flexibly and avoids the misspecification bias that parametric specifications introduce. However, ML estimation leads to regularization and model selection biases that standard inference procedures do not account for, threatening the validity of inference when ML-generated outcomes are used as dependent variables. The empirical literature exploiting ML's flexibility typically plugs ML-generated outcomes directly into standard estimators without correcting for regularization bias, yielding biased point estimates and confidence intervals that fail to achieve nominal coverage. Relying instead on parametric specifications avoids regularization bias at the cost of misspecification bias that propagates through all subsequent stages. This trade-off motivates the development of a framework that preserves the flexibility of machine learning while delivering valid inference despite the regularization bias it introduces.\par
This paper develops a framework for locally robust inference with outcomes predicted by machine learning. We consider a three-step setting in which the dependent variable is generated in a first step via machine learning regression of an observed outcome on a vector of covariates. In a second step, a nuisance function is constructed using possibly another machine learning projection of the generated outcome. Finally, the parameter of interest is identified in a third step through a Generalized Method of Moments (GMM) criterion that integrates the outputs of the first two steps with the observed data. By accommodating ML-generated outcomes, our framework enables researchers to exploit the flexibility of machine learning while delivering valid inference.\par 
A closely related literature studies inference when machine-learning predictions are used as outcomes in subsequent statistical analysis. \cite{wang2020postprediction} show that treating predicted outcomes as observed can distort downstream inference and propose post-prediction corrections based on separate training, testing, and validation samples. \cite{angelopoulos2023prediction} develop prediction-powered inference, which combines labeled observations with predictions from an arbitrary machine-learning system to conduct inference on features of the underlying observed outcome distribution. These approaches establish that prediction error must be accounted for in downstream inference. Our problem is complementary. The population prediction function $g_0$ is itself an unknown nuisance entering the parameter of interest, and it may also generate a subsequent nuisance $h_0(g_0)$. We derive orthogonal GMM moments that remove the first-order effects of estimating these objects by machine learning. In particular, ordinary sample splitting or cross-fitting makes predictions out of sample but does not by itself eliminate the first-order effect of estimating $g_0$ on a downstream parameter defined through that function.\par
Our paper builds on Neyman-orthogonal inference \citep{chernozhukov2018double,chernozhukov2022locally} and the three-step generated-regressor analysis of \cite{escanciano2023automatic}. \cite{chernozhukov2023automatic} develop recursive Riesz corrections for nested conditional-mean and instrumental-variable functionals. We develop a generated-outcome GMM framework that separates the direct effect of the first-step prediction from its indirect effect through a downstream nuisance, making the role of downstream local robustness explicit. The framework also accommodates nuisance functions defined by projection onto restricted linear spaces, without requiring the projection residuals to have zero conditional mean. The constructed-instrument application to productivity persistence relies on this feature.\par
Our proposal consists of constructing a locally robust/debiased moment condition that augments the original identifying moment with the first- and second-step influence functions. In doing so, we account for both the effect of the ML-generated outcome and the nonparametric nuisance estimation on inference. Specifically, after correcting for the second step, the first-step correction is obtained by differentiating the augmented moment with respect to the first step, holding the second-step nuisance fixed. The resulting moment condition satisfies Neyman orthogonality with respect to both nuisance parameters, ensuring inference on the parameter of interest is locally insensitive to the regularization and model selection biases introduced by their ML estimation.\par 
We illustrate our framework through three examples. First, we extend the partially linear model to an ML-predicted dependent variable, which enters the identifying moment directly and generally requires a first-step correction. Second, we study the persistence of unobserved productivity with an unrestricted production function. Rather than assuming a valid instrument, we construct instruments by orthogonalizing period-1 productivity signals with respect to additive functions of period-1 and period-2 inputs, characterize when relevant instruments exist, and account for the ML estimation of both residualized productivity and the constructed instrument. Finally, we propose a life-cycle correction for the measurement of inequality of opportunity, in which incomes are adjusted with type-specific age profiles estimated by ML. The generated outcome then enters inequality of opportunity both directly, through total inequality, and indirectly, through the type averages.\par\par 
We establish consistency and asymptotic normality of the proposed locally robust estimator under mild conditions. We study its finite-sample behavior through Monte Carlo simulations for the partially linear model. In our simulations, the locally robust estimator has negligible bias that vanishes as the sample size increases and coverage rates close to nominal levels. In contrast, the naive plug-in estimator, which treats the ML-generated outcome as observed and applies no correction for regularization bias, exhibits sizable bias and coverage rates far below nominal, confirming that ignoring the generated nature of the outcome leads to invalid inference.\par   
We apply our framework to measure inequality of opportunity in Spain, using the 2023 wave of the Encuesta de Condiciones de Vida. The survey records a single year of income for individuals observed at different ages, so measured inequality of opportunity is particularly exposed to life-cycle bias. We correct incomes with type-specific age profiles estimated by machine learning, evaluated over the working life. Using the variance of logs and the mean log deviation, we find that circumstances account for about 9\% of income inequality once corrected income is treated as a generated outcome, 10\% to 14\% less than with raw income, whereas treating corrected income as observed overstates this share by up to 15\%.\par
The remainder of this paper proceeds as follows. Section~\ref{sec:set} presents the three-step setting and the general construction of locally robust moments in the presence of a generated dependent variable. Section~\ref{sec:illustrate} illustrates the framework through three examples. Section~\ref{sec:estim} develops the debiased GMM estimator, establishes its consistency and asymptotic normality, and examines its finite-sample performance through Monte Carlo simulations. Section~\ref{sec:empirical} applies the method to the measurement of inequality of opportunity in Spain, and Section~\ref{sec:conc} concludes. Appendix~\ref{sec:proofs} contains the proofs.
\section{A General Framework for Valid Inference with ML Generated Outcomes}\label{sec:set}
\subsection{Three-Step Setting}
We observe data $W = (Y, X)$ with cumulative distribution function (cdf) $F_0$, where $Y$ is an observed outcome and $X$ collects all available covariates. The parameter of interest $\theta_0$ is identified by a moment condition that depends on an outcome $\tilde{Y}$ not directly observed, but instead generated from $Y$ through a regression function $g_0$ of some components of $X$, estimated via machine learning. We develop locally robust inference for $\theta_0$ that remains valid despite the regularization and model selection biases introduced by ML estimation of $g_0$. We consider the case where the generated outcome is one-dimensional, although our framework naturally extends to the multivariate case by applying the construction to each component.\par
We describe our general three-step setting as follows:\par \bigskip 
\textbf{First step.} 
Let $X_1$ denote some components of the available covariates $X$, used to construct the generated outcome. There is a first step $g_0(X_1)$ solving the moment restrictions
\begin{align}\label{eq:gen}
\mathbb{E}[\delta_1(X_1)\left(Y - g_0(X_1)\right)] = 0 \quad \text{for all } \delta_1 \in \Delta_1,\end{align}
where $Y - g_0(X_1)$ is the prediction error for $g_0 \in \Delta_1$, and $\Delta_1$ is a linear and closed subspace of $L_2(X_1)$. Henceforth, for a generic random variable $U$, we denote by $L_2(U)$ the Hilbert space of square-integrable functions of $U$, i.e.\ $g \in L_2(U)$ iff $\mathbb{E}[g^2(U)] < \infty$. If $\Delta_1 = L_2(X_1)$, then $g_0(X_1) = \mathbb{E}[Y|X_1]$, as when the researcher predicts an economic outcome from satellite imagery or constructs a counterfactual continuation value in a dynamic treatment regime. Since $\text{dim}(X_1)$ is allowed to be high, this nonparametric first step may not be feasible to implement directly. We can then consider a high-dimensional additive regression model with the same error with $\Delta_1 = \sum_{j=1}^{\text{dim}(X_1)}\Delta_{1,j}$, where $\Delta_{1,j}$ is a subset of $L_2(X_{1j})$ for $X_{1j}$ the $j$-th component of $X_1$. When $\Delta_1$ is the mean square limit of linear combinations $\sum_{k=1}^{K}\beta_{0k}c_k(X_1)$ for $K \in \mathbb{N}$, a sequence of real numbers $(\beta_{0k})_{k=1}^{\infty}$, and a dictionary $(c_k)_{k=1}^{\infty}$ of functions in $L_2(X_1)$, this setting covers high-dimensional linear regression, commonly estimated via Lasso or other ML methods.\par
The first step $g_0$ in \eqref{eq:gen} is used to construct the generated outcome
\begin{align}\label{eq:varphi} \tilde{Y} \equiv \tilde{Y}(g_0) \equiv \varphi(W, g_0), \end{align}
where $\varphi$ is a known mapping from observed variables $W$ and the unknown function $g_0$ to the generated outcome $\tilde{Y}$. In the leading case $\varphi(W, g_0) = g_0(X_1)$, the generated outcome is directly the ML prediction of $Y$ given $X_1$. Alternatively, $\varphi(W, g_0) = Y - g_0(X_1)$ covers settings where the outcome of interest is  residualized of confounding variation, such as income or consumption purged of life-cycle variation (See Section~\ref{sec:iop}).\par
The framework also covers designs in which $Y$ is observed only in part of the sample. Let $R=1$ indicate the units for which $Y$ is observed, take $W$ to be the pooled sample with observed outcome $RY$, and let $X_1=(V,R)$, where $V$ collects the predictors. The first step then gives $g_0(v,1)=\mathbb{E}[Y\mid V=v,R=1]$, and the generated outcome $\varphi(W,g_0)=g_0(V,1)$ evaluates the prediction at $R=1$ for every unit. In the surrogate design of \citet{athey2025surrogate}, $R=1$ indicates the observational sample, $V$ collects the surrogates and pre-treatment covariates, and $g_0(V,1)$ is the surrogate index. The first-step correction is then supported on the units with $R=1$, and identification requires, in addition, assumptions specific to each design, such as surrogacy, comparability and overlap across samples.\par\bigskip
\textbf{Second step.} 
Let $S \equiv S(g_0)$ denote some components of $(\tilde{Y}(g_0), W)$, and $X_2$ denote some components of $X$ onto which $S$ is projected. The second step is an orthogonal projection of $S$ on $X_2$, identified by
\begin{align}
\label{eq:orth2} \mathbb{E}[\delta_2(X_2)(S(g_0) - h_0(X_2))] = 0 \quad \text{for all } \delta_2 \in \Delta_2, \end{align}
where, if $S(g_0)$ is $d_S$-dimensional, $\Delta_2$ denotes the corresponding product Hilbert space $\Delta_2^{(1)}\times\cdots\times\Delta_2^{(d_S)}$, with each $\Delta_2^{(j)}$ a closed linear subspace of $L_2(X_2)$. In the scalar case, this reduces to a single closed linear subspace of $L_2(X_2)$. Here, $S(g_0)$ may be the generated outcome itself, a vector containing it alongside observed regressors, or a transformed version of it (as when partialling out confounders from an ML-predicted dependent variable or projecting a counterfactual outcome onto period-1 state variables). This dependence of $h_0$ on $g_0$ through $S(g_0)$ gives the problem a compositional structure that we exploit explicitly in the construction of locally robust moments and in the corresponding asymptotic analysis. 
\par \bigskip
\textbf{Third step.} 
Let $\theta_0 \in \Theta \subseteq \mathbb{R}^{p}$ denote the $p$-dimensional parameter of interest, identified by $q \geq p$ moment conditions. The moment function $m\colon \mathbb{R}^{\text{dim}(W)}\times \Delta_1 \times \Delta_2 \times \Theta \rightarrow \mathbb{R}^{q}$ identifies $\theta_0$ through the GMM moment condition
\begin{align}\label{eq:main} \mathbb{E}[m(W,g_{0},h_{0},\theta_{0})]=0,\end{align}
where $\theta_0$ is assumed to be the unique solution to $\mathbb{E}[m(W,g_{0},h_{0},\theta)]=0$ over $\theta \in \Theta$. The moment $m$ may depend on $g_0$ directly, as in the partially linear model where the dependent variable is an ML prediction, or only through $h_0$, as in dynamic treatment effects where the parameter of interest is identified by a marginal structural moment condition.\par 
\subsection{Locally Robust Moment Construction}
The moment in equation (\ref{eq:main}) identifies the parameter $\theta_0$ as a function of the observed data $W$ and the nuisance parameters $g_0$ and $h_0$. Accordingly, we can consider the nuisance parameter $\gamma(F)\equiv\left(g(F), h(F,g(F))\right)$, where $h(F,g)$ denotes the second-step nuisance when $W$ follows the distribution $F$ and the first-step is fixed at $g$, so that the moment identifying the parameter of interest $\theta_0$ in equation (\ref{eq:main}) can be expressed as
\begin{align}
\label{eq:main2}\bar{m}(\gamma_0,\theta_0)\equiv\mathbb{E}[m(W,\gamma_0,\theta_0)]=0,
\end{align}
where $\bar{m}(\gamma,\theta)$ denotes the expected moment function, $\mathbb{E}\left[\cdot\right]$ is the expectation under $F_0$, $g_0\equiv g(F_0)$, $h_0\equiv h(F_0,g_0)$, and $\gamma_0\equiv(g_0, h_0)$ is the probability limit (plim) of $\hat{\gamma}$ under $F_0$.\par 
The challenge of estimating $\theta_0$ using equation (\ref{eq:main2}) lies in the sensitivity of the identifying moment to estimation of $\gamma_0$. Formally, for an alternative distribution function $H$ and a regular path $F_\tau=(1-\tau)F_0+\tau H$ for $\tau \in [0,1]$, such that $\gamma(F_\tau)\equiv(g(F_\tau), h(F_\tau, g(F_\tau)))$ exists for $\tau$ small enough, the identifying moment is not generally locally robust to estimation of $\gamma_0$, i.e., 
\begin{align*}\frac{d}{d\tau}\bar{m}(\gamma(F_\tau),\theta)\neq 0,\end{align*}  
where, henceforth, $d/d\tau$ is the derivative with respect to $\tau$ from the right and evaluated at $\tau=0$.\par 
This motivates constructing a moment condition whose expectation is insensitive to local perturbations of $\gamma$ around $\gamma_0$, that is, one whose derivative with respect to $\tau$ vanishes at $\tau = 0$, a property known as Neyman orthogonality \citep{neyman1959optimal,chernozhukov2018double, chernozhukov2022locally}, or equivalently, local robustness with respect to $(g_0, h_0)$.\par 
We derive the following standard form for the debiased/locally robust moment function:
\begin{align}\label{eq:debiased}
\psi(w, g_0, h_0, \alpha_0, \theta)= m(w, g_0, h_0, \theta)+ \underbrace{\alpha_{01}(x_1;\theta) \cdot (y - g_0(x_1))}_{\equiv \phi_1(w,g_0,\alpha_{01},\theta)} + \underbrace{\alpha_{02}(x_2;\theta)' \left(s - h_0(x_2)\right)}_{\equiv \phi_2(w,h_0,\alpha_{02},\theta)}, \end{align} 
where $\alpha_0\equiv(\alpha_{01},\alpha_{02})$ denotes the Riesz representers and \[ \phi\equiv\phi_1+\phi_2 \] denotes the total influence-function correction, with $\phi_1$ and $\phi_2$ the first- and second-step corrections, respectively. The first-step influence function corresponds to $\phi_1$ in equation (\ref{eq:debiased}) and captures the effect of the generated variable on the identifying moment. In general, it is nonzero, implying that inference procedures that ignore generated outcomes are invalid. Importantly, $\phi_1$ depends on the first-step Riesz representer $\alpha_{01}$ and the corresponding identifying error $\epsilon(w, g_0)=y - g_0(x_1)$. The second-step influence function, $\phi_2$, accounts for the effect of estimating $h_0$ on the moment condition and similarly depends on $\alpha_{02}$. When there are finitely many first-step regressions, as in the production-function example, we interpret $\Delta_1$ as the corresponding product Hilbert space. The first-step correction is then the sum of the componentwise corrections, which follows directly from linearity of the Hadamard derivative and the Riesz representation on the product space. \par 
To construct locally robust/debiased estimators, we show that Neyman orthogonal moments are constructed by sequentially adding to the original identifying moment the second- and first-step IFs. This sequential construction exploits the idea of downstream local robustness of \cite{escanciano2023automatic}: after correcting for $h$, the first-step IF is obtained by differentiating the augmented moment with respect to $g$, holding $h$ fixed. To fix ideas, notice that the impact of $\gamma(F_\tau)$ on $\bar{m}$ is given by:
\begin{align*}
\frac{d}{d\tau}\bar{m}(\gamma(F_\tau),\theta) &= \frac{d}{d\tau}\bar{m}(g(F_\tau), h(F_0, g(F_\tau)),\theta) + \frac{d}{d\tau}\bar{m}(g_0, h(F_\tau, g_0),\theta) \nonumber
\\&= \underbrace{\frac{d}{d\tau}\bar{m}(g(F_\tau), h_0,\theta)}_{\text{direct effect of }g} + \underbrace{\frac{d}{d\tau}\bar{m}(g_0, h(F_0, g(F_\tau)),\theta)}_{\text{indirect effect of }g\text{ through }h} + \underbrace{\frac{d}{d\tau}\bar{m}(g_0, h(F_\tau, g_0),\theta)}_{\text{direct effect of }h} \nonumber
\\&= \frac{d}{d\tau}\bar{m}(g(F_\tau), h_0,\theta) + \frac{d}{dh}\bar{m}(g_0, h_0,\theta)\cdot\left[\frac{d}{d\tau}h(F_0, g(F_\tau)) + \frac{d}{d\tau}h(F_\tau, g_0)\right],
\end{align*} 
where the last equality follows by the chain rule, under Hadamard differentiability of $\bar{m}(g,h,\theta)$ with respect to $h$ and of $h(F_0,g)$ with respect to $g$; see Proposition~3.1 of \citet{escanciano2023automatic}. The first equality requires, in addition, that $\tau\mapsto g(F_\tau)$ be differentiable and that the remainder $h(F_\tau,g(F_\tau))-h(F_0,g(F_\tau))-h(F_\tau,g_0)+h_0$ be $o(\tau)$, which holds under joint Hadamard differentiability of $(F,g)\mapsto h(F,g)$.\par 
The last display highlights that if the moment condition is locally robust to $h$, meaning $$\frac{d}{dh}\bar{m}(g_0, h_0,\theta)=0,$$ then there is no indirect effect of $g$ through $h$. This is the idea of the downstream local robustness in \cite{escanciano2023automatic} applied to generated outcomes. Exploiting this insight, we proceed sequentially: first, we augment $m$ with the influence function  of $h$,
\begin{align*} m^*(w, g_0, h_0, \alpha_{02}, \theta)\equiv m(w, g_0, h_0,\theta)+\phi_2(w,h_0,\alpha_{02},\theta), \end{align*}
and then we augment $m^*$ with the influence function of $g$, yielding
\begin{align*} \psi(w, g_0, h_0, \alpha_0, \theta)&\equiv m^*(w, g_0, h_0, \alpha_{02}, \theta)+\phi_1(w,g_0,\alpha_{01},\theta)
\\&=m(w, g_0, h_0,\theta)+\phi_1(w,g_0,\alpha_{01},\theta)+\phi_2(w,h_0,\alpha_{02},\theta).
\end{align*}
Notice that the locally robust moment depends on the Riesz representers $\alpha_0 \equiv (\alpha_{01}, \alpha_{02})$, which are identified by the orthogonality conditions (with the notation $h_0(g)\equiv h(F_0,g)$)
\begin{align*}\mathbb{E}[\alpha_{01}(X_1;\theta)\delta_1(X_1)]&= \frac{d}{d\tau} \mathbb{E}\!\left[ m\left(W, g_0+\tau\delta_1, h_0(g_0+\tau\delta_1), \theta\right) \right]\bigg|_{\tau=0} \quad \text{for all } \delta_1 \in \Delta_1,\\
\mathbb{E}[\alpha_{02}(X_2;\theta)'\delta_2(X_2)]&= \frac{d}{d\tau}\mathbb{E}\!\left[ m\left(W,g_0,h_0(g_0)+\tau\delta_2,\theta\right) \right]\bigg|_{\tau=0} \quad \text{for all } \delta_2 \in \Delta_2.\end{align*}
Here $\alpha_{01}$ represents the Riesz representer of the total Gateaux derivative of the expected moment with respect to the first-step nuisance $g$, including both the direct effect of $g$ on the moment and the indirect effect through the induced change in $h_0(g)$. By contrast, $\alpha_{02}$ represents the Riesz representer of the Gateaux derivative with respect to the second-step nuisance $h$, holding $g_0$ fixed. Equivalently, since $m^*$ is locally robust to $h$, $\alpha_{01}$ represents the derivative of $\mathbb{E}[m^*]$ with respect to $g$, holding $h$ fixed at $h_0$.\par
The following assumption imposes pathwise differentiability and Riesz representability conditions ensuring that the orthogonal corrections are well-defined. We begin by recalling the notion of Hadamard tangential differentiability.
\begin{definition}
Let $\mathbb{D}$ and $\mathbb{E}$ be normed vector spaces, let $\mathbb{D}_0\subseteq\mathbb{D}$, and let $\phi:\mathbb{D}_\phi\subseteq\mathbb{D}\to\mathbb{E}$. The map $\phi$ is Hadamard differentiable at $\eta\in\mathbb{D}_\phi$ tangentially to $\mathbb{D}_0$ if there exists a continuous linear map $D\phi(\eta):\mathbb{D}_0\to\mathbb{E}$ such that
\[\frac{ \phi(\eta+t_n h_n)-\phi(\eta)}{t_n} \to D\phi(\eta)[h]\]
for every sequence $t_n\downarrow0$ and every sequence $h_n\to h\in\mathbb{D}_0$ satisfying $\eta+t_n h_n\in\mathbb{D}_\phi$ for all sufficiently large $n$. The map $D\phi(\eta)$ is called the Hadamard derivative of $\phi$ at $\eta$.
\end{definition}
\begin{assumption} \label{ass:lr}
For each $\theta\in\Theta$, let 
\[\bar m(g,h,\theta)\equiv \mathbb{E}[m(W,g,h,\theta)]\]
and define $h_0(g)\equiv h(F_0,g)$, with $h_0\equiv h_0(g_0)$. The map \[ (g,h)\mapsto \bar m(g,h,\theta) \] is Hadamard differentiable at $(g_0,h_0)$ tangentially to $\Delta_1\times\Delta_2$. Moreover, the generated outcome map $g\mapsto S(g)$, viewed as a map from $\Delta_1$ into the appropriate $L_2(F_0)$ product space, is Hadamard differentiable at $g_0$ tangentially to $\Delta_1$.
\end{assumption}
Under Assumption \ref{ass:lr}, the second-step map $g\mapsto h_0(g)$ is also Hadamard differentiable. Indeed, since $h_0(g)$ is the $L_2$ projection of $S(g)$ onto $\Delta_2$, its derivative is the corresponding projection of $D_gS(g_0)[\delta_1]$. Thus, denoting
\[\dot h_g(\delta_1)\equiv D_gh_0(g_0)[\delta_1], \]
the derivative $\dot h_g(\delta_1)\in\Delta_2$ is characterized by
\[ \mathbb{E}\left[ \delta_2(X_2) \left\{ D_gS(g_0)[\delta_1] - \dot h_g(\delta_1)(X_2) \right\} \right]=0 \quad \text{for all } \delta_2\in\Delta_2. \]
Consequently, by the chain rule, the composite map $g\mapsto \bar m(g,h_0(g),\theta)$ is Hadamard differentiable at $g_0$.\par 
Define the partial derivative functionals
\[\dot m_g(\delta_1;\theta)\equiv D_g\bar m(g_0,h_0,\theta)[\delta_1], \qquad \dot m_h(\delta_2;\theta) \equiv D_h\bar m(g_0,h_0,\theta)[\delta_2],\]
and the total first-step derivative
\[\dot m_{g,\mathrm{tot}}(\delta_1;\theta) \equiv \dot m_g(\delta_1;\theta) + \dot m_h(\dot h_g(\delta_1);\theta). \]
Equivalently,
\[ \dot m_{g,\mathrm{tot}}(\delta_1;\theta) = \frac{d}{d\tau} \mathbb{E}\!\left[ m\{W,g_0+\tau\delta_1,h_0(g_0+\tau\delta_1),\theta\} \right]\bigg|_{\tau=0}. \]
When $q>1$, the following Riesz representations are understood componentwise, so that each component of the $q$-dimensional moment has its own first- and second-step representers. The Riesz representers are defined by
\begin{align} \mathbb{E}[\alpha_{01}(X_1;\theta)\delta_1(X_1)] &= \dot m_{g,\mathrm{tot}}(\delta_1;\theta), \quad \text{for all } \delta_1\in\Delta_1, \label{eq:riesz1}
\\ \mathbb{E}[\alpha_{02}(X_2;\theta)'\delta_2(X_2)] &= \dot m_h(\delta_2;\theta), \quad \text{for all } \delta_2\in\Delta_2. \label{eq:riesz2} \end{align}

\begin{theorem}
\label{thm:lr}
Suppose Assumption \ref{ass:lr} holds. Then there exist unique Riesz representers $\alpha_0=(\alpha_{01},\alpha_{02})$, with $\alpha_{01}\in\Delta_1^q$ and $\alpha_{02}\in\Delta_2^q$, satisfying \eqref{eq:riesz1}--\eqref{eq:riesz2} componentwise. Define
\[ \psi(w,g,h,\alpha_0,\theta) = m(w,g,h,\theta) + \alpha_{01}(x_1;\theta)(y-g(x_1)) + \alpha_{02}(x_2;\theta)'(s(g)-h(x_2)), \] 
where $s(g)$ denotes the realization of $S(g)$. Then
\[\mathbb{E}[\psi(W,g_0,h_0,\alpha_0,\theta_0)]=0\]
whenever
\[\mathbb{E}[m(W,g_0,h_0,\theta_0)]=0. \]
Moreover, $\psi$ is Neyman orthogonal: for every regular path
\[ g_\tau=g_0+\tau\delta_1+o(\tau), \qquad h_\tau=h_0+\tau\delta_2+o(\tau),\]
with $\delta_1\in\Delta_1$ and $\delta_2\in\Delta_2$,
\[ \frac{d}{d\tau} \mathbb{E}\!\left[ \psi(W,g_\tau,h_\tau,\alpha_0,\theta)\right]\bigg|_{\tau=0} = 0.\]
\end{theorem}
\begin{remark}
\label{rem:linear}
In the leading case where $\Delta_2=L_2(X_2)$, the derivative of the second-step projection satisfies
\[D_gh_0(g_0)[\delta_1](X_2) = \mathbb{E}[D_gS(g_0)[\delta_1]\mid X_2].\]
Let $a_g(X_1;\theta)$ denote the Riesz representer of the direct partial derivative $\dot m_g(\delta_1;\theta)$. Since $\alpha_{01}$ represents the total derivative
\[\dot m_{g,\mathrm{tot}}(\delta_1;\theta)=\dot m_g(\delta_1;\theta)+\dot m_h(\dot h_g(\delta_1);\theta), \]
it follows that
\[ \alpha_{01}(X_1;\theta)=a_g(X_1;\theta) +\Pi_{\Delta_1}\mathbb{E}[\alpha_{02}(X_2;\theta)\mid X_1]\]
when $S(g)=g(X_1)$, while
\[ \alpha_{01}(X_1;\theta) = a_g(X_1;\theta) - \Pi_{\Delta_1}\mathbb{E}[\alpha_{02}(X_2;\theta)\mid X_1]\]
when $S(g)=Y-g(X_1)$, where $\Pi_{\Delta_1}$ denotes the $L_2$ projection onto $\Delta_1$, applied componentwise, which is the identity when $\Delta_1=L_2(X_1)$. In particular, if $\alpha_{02}=0$, then $\alpha_{01}=a_g(X_1;\theta)$, corresponding to the absence of an indirect effect through the second-step nuisance, a special case of downstream local robustness.
\end{remark}
\begin{remark}
For simplicity of notation, we suppress the dependence of the Riesz representers $\alpha_{01}(\cdot;\theta)$ and $\alpha_{02}(\cdot;\theta)$, and other nuisance objects on $\theta$ whenever no confusion may arise. \end{remark}
\section{Three Leading Examples}\label{sec:illustrate}
This section illustrates our general framework through three leading cases. First, we extend the partially linear model to allow for an ML-predicted dependent variable, where the generated outcome enters the moment condition directly, and a nonzero first-step correction is required. Second, we study the estimation of the persistence of unobserved productivity with an unrestricted production function. This application features two first-step regressions: one generates residualized productivity and enters the identifying moment through a downstream conditional expectation, while the other generates a constructed instrument and enters the identifying moment directly. Finally, we propose a life-cycle correction for the measurement of inequality of opportunity, in which a single first-step regression enters the identifying moment both directly, through total inequality, and indirectly, through the type averages.
\subsection{The Partially Linear Model}\label{subsec:plm}
In this context, we observe $W = (Y, X)$, where $X$ includes a vector $X_1$ used to predict $Y$ in a first step, a vector of controls $X_2$ onto which the generated outcome is projected in a second step, and a low-dimensional vector of regressors of interest $X_3$ whose effect on the ML prediction is the parameter of interest in a third step. Although $X_2$ and $X_3$ are disjoint, we allow $X_1$ to overlap with either or contain additional covariates in $X$. For instance, $X_3$ may be a policy variable or a socioeconomic characteristic of interest, $X_2$ demographic and geographic controls, and $X_1$ may collect satellite-derived predictors of an economic outcome unavailable at scale, potentially augmented by $X_2$.\par
Let $\Delta_1\subseteq L_2(X_1)$ be a specified closed linear prediction space, and let $\Pi_{\Delta_1}$ denote the $L_2(F_0)$ orthogonal projection onto that space. The generated outcome is
\[\tilde Y=g_0(X_1),\qquad g_0=\Pi_{\Delta_1}Y.\]
Thus, $g_0$ is the best mean-square predictor within $\Delta_1$, which need not equal $\mathbb E[Y\mid X_1]$. We estimate $g_0$ using machine learning and use its prediction as the dependent variable in the model
\begin{align}\label{eq:plm} \tilde{Y} = X_3'\theta_0 + r_0(X_2) + U,\qquad\mathbb{E}[U|X_2,X_3] = 0,\end{align}
which identifies $\theta_0$ as the partially linear projection coefficient of the ML prediction on $X_3$ controlling for $X_2$. \par
Following \cite{robinson1988root}, the identifying moment condition for $\theta_0$ is
\begin{align}\label{eq:robinsmoment} \mathbb{E}\left[ \left( \tilde{Y} - h_{0Y}(X_2) - \theta_0'(X_3-h_{0X}(X_2)) \right) (X_3-h_{0X}(X_2)) \right] = 0, \end{align}
where
\[ h_{0Y}(X_2)=\mathbb{E}[\tilde{Y}|X_2], \qquad h_{0X}(X_2)=\mathbb{E}[X_3|X_2].\]
However, since $\tilde{Y}$ is itself estimated by machine learning, this moment condition does not account for the regularization bias introduced in the first step. To construct the locally robust moment, we start by mapping the problem into our three-step framework.
\paragraph{Three-step representation.}
This problem fits our framework as follows.\par
\bigskip \noindent
\textbf{First step.}
The first-step function $g_0\in\Delta_1$ is identified by
\[ \mathbb E\left[ \delta_1(X_1)\{Y-g_0(X_1)\} \right]=0 \quad \text{for all }\delta_1\in\Delta_1. \]
The generated outcome is
\[\tilde Y=\varphi(W,g_0)=g_0(X_1).\]
\bigskip \noindent
\textbf{Second step.}
Let $X_2$ denote the controls onto which the generated outcome and the regressors of interest are projected, and define
\[ S(g_0)\equiv(\tilde{Y}(g_0),X_3)', \qquad h_0(X_2) = \mathbb{E}[S(g_0)|X_2] = (h_{0Y}(X_2),h_{0X}(X_2))'.\]

\bigskip \noindent
\textbf{Third step.}
The parameter $\theta_0$ is identified by the moment condition
\[ \mathbb{E}[m(W,g_0,h_0,\theta_0)]= 0,\]
where
\[m(W,g,h,\theta)=\big(g(X_1)-h_Y(X_2)-\theta'(X_3-h_X(X_2))\big)(X_3-h_X(X_2)),\]
with $h=(h_Y,h_X)'$.

The following proposition derives the locally robust moment for$\theta_0$.
\begin{proposition}
\label{prop:plm}
Let $g_0=\Pi_{\Delta_1}Y$, and define
\[h_{0Y}(X_2)=\mathbb E[g_0(X_1)\mid X_2], \qquad h_{0X}(X_2)=\mathbb E[X_3\mid X_2], \qquad \nu=X_3-h_{0X}(X_2). \]
Suppose $B=\mathbb E[\nu\nu']$ is nonsingular and let
\[ \theta_0 = B^{-1}\mathbb E[ \{g_0(X_1)-h_{0Y}(X_2)\}\nu ].
\]
The first- and second-step Riesz representers are
\[ \alpha_{01}=\Pi_{\Delta_1}\nu, \qquad \alpha_{02}=0, \]
where the projection is applied component-wise. A locally robust moment for $\theta_0$ is
\begin{align} \psi(W,g_0,h_0,\alpha_0,\theta)
&=\{g_0(X_1)-h_{0Y}(X_2)-\theta'\nu\}\nu \nonumber\\ &\quad +\alpha_{01}(X_1)\{Y-g_0(X_1)\}. \label{eq:lr_plm_general_main} \end{align}

If $\Delta_1=L_2(X_1)$, then
\[g_0(X_1)=\mathbb E[Y\mid X_1],\qquad \alpha_{01}(X_1)=\mathbb E[\nu\mid X_1]. \]
If, additionally, $X_2\subseteq X_1$ and $X_3\subseteq X_1$, then $\alpha_{01}=\nu$ and $h_{0Y}(X_2)=\mathbb E[Y\mid X_2]$, so
\begin{align}\psi(W,h_0,\theta)=\{Y-h_{0Y}(X_2)-\theta'\nu\}\nu.\label{eq:lr_plm_main} \end{align}
\end{proposition}
\begin{remark}[Partially linear projection]
\label{rem:plm_projection}
The conditional mean restriction in \eqref{eq:plm} is stronger than needed for the moment construction. Let
\[ \nu\equiv X_3-h_{0X}(X_2),\qquad B\equiv\mathbb{E}[\nu\nu'],\]
and define the partially linear projection coefficient
\[\theta_0= B^{-1}\mathbb{E}\big[ (\tilde Y-h_{0Y}(X_2))\nu\big].\]
Setting
\[U=\tilde Y-h_{0Y}(X_2)-\theta_0'\nu\]
gives
\[\mathbb{E}[U\mid X_2]=0,\qquad\mathbb{E}[U\nu]=0.\]
These conditions are sufficient for the orthogonality calculations in Proposition~\ref{prop:plm}. Hence the locally robust moments \eqref{eq:lr_plm_general_main}--\eqref{eq:lr_plm_main}, and the asymptotic results below, remain valid when $\theta_0$ is interpreted as this partially linear projection coefficient.
\end{remark}
A policy-relevant example of this setting arises in algorithmic credit scoring. Lenders increasingly rely on machine-learning models of default risk that exclude protected characteristics such as race, as fair lending rules typically require, although the remaining borrower information may still proxy for them \citep{fuster2022predictably,hurlin2026fairness}. Whether such a race-blind score treats comparable borrowers differently is then a natural parameter of interest, and it depends on a generated outcome. When the score is the conditional default probability, $\Delta_1=L_2(X_1)$, and the comparison variables satisfy $X_2\subseteq X_1$, Proposition~\ref{prop:plm}, with group membership $A$ in the role of $X_3$, gives $\alpha_{01}(X_1)=\mathbb E[A\mid X_1]-\mathbb E[A\mid X_2]$, so the first-step correction is governed by how much the credit model's information predicts group membership beyond the comparison variables. The same structure arises whenever disparities are measured in algorithmic scores, as in insurance pricing, hiring or the targeting of public programs.\par
The cancellation in \eqref{eq:lr_plm_main} is a special case. For a restricted first-step prediction space,
\[\mathbb E[Y\mid X_2]-h_{0Y}(X_2)= \mathbb E[Y-g_0(X_1)\mid X_2],\]
which need not vanish even when $X_2\subseteq X_1$. Thus, in general, $h_{0Y}$ must be estimated from the generated outcome rather than by regressing $Y$ directly on $X_2$. The identity $h_{0Y}(X_2)=\mathbb E[Y\mid X_2]$ holds under the additional condition $\mathbb E[Y-g_0(X_1)\mid X_2]=0$, for example when $L_2(X_2)\subseteq\Delta_1$. We do not impose this condition in the general PLM formulation.\par 
Although the second-step correction is zero, the second-step nuisance remains generated: its estimation depends on the first-step prediction. Orthogonality removes its first-order effect on the expected moment, not its role in constructing the estimator.
\begin{remark}[Outcomes residualized of confounding variation] \label{rem:resid}
This PLM setting extends directly to a common practice in empirical work: residualizing an observed outcome of confounding variation before using it in subsequent analysis. In the measurement of inequality of opportunity, income is purged of life-cycle or cohort variation before its remaining dispersion is attributed to circumstances beyond individual control \citep{brunori2025inequality}; in partial insurance models, consumption is purged of predictable components before household smoothing is studied \citep{blundell2008consumption}. The
object entering the subsequent analysis is then a generated outcome, corresponding to the choice $\tilde{Y}\equiv\varphi(W,g_0)=Y-g_0(X_1)$.\par 
Robinson's moment remains orthogonal with respect to $h_0$, so $\alpha_{02}=0$. Perturbing $g$ enters with the opposite sign to the predicted-outcome case, giving \[\alpha_{01}=-\Pi_{\Delta_1}\nu,\qquad\nu=X_3-h_{0X}(X_2),\] where the projection is applied componentwise. The locally robust moment is therefore \[\psi=\left(\tilde Y-h_{0Y}(X_2)-\theta'\nu\right)\nu+\alpha_{01}(X_1)\left(Y-g_0(X_1)\right),\] where \[h_{0Y}(X_2)=\mathbb E[\tilde Y\mid X_2], \qquad\tilde Y=Y-g_0(X_1).\]
Thus, the first-step identifying error coincides with the generated outcome itself. In the conditional-mean case $\Delta_1=L_2(X_1)$, the representer simplifies to
\[\alpha_{01}(X_1)=-\mathbb E[\nu\mid X_1].\]
The same construction applies when $\theta_0$ is interpreted as
the partially linear projection coefficient in
Remark~\ref{rem:plm_projection}.
\end{remark}

\subsection{Productivity Persistence with Nonparametric Production Functions}

A relevant application in economics where generated outcomes arise
is the study of productivity persistence, as productivity is not
directly observed but is recovered using an estimated production
function. Our framework accommodates structural environments in which the
production function is left unrestricted. We consider identification
and inference on the persistence of latent productivity. Rather than
requiring an externally specified instrument, we construct valid
instruments from observable period-1 productivity signals using
orthogonal projections. This construction separates instrument validity
from relevance and fits directly into our general locally robust
framework.

Productivity is a key driver of firm performance, yet its latent nature
requires inferring its dynamics from data on inputs and outputs. We
observe
\[
W=\big((Y_t,K_t,L_t,I_t)_{t=1,2}\big),
\]
where $Y_t$ is output, $(K_t,L_t)$ are capital and labor inputs, and
$I_t$ is investment, all in logs. Our parameter of interest is the
persistence of latent productivity, denoted by $\theta_0$.

\paragraph{Structural model and timing.}
Assume output satisfies
\begin{align}
Y_t
=
\tilde g_{0t}(K_t,L_t)+\omega_t+U_t,
\qquad t=1,2,
\label{eq:prod_struct}
\end{align}
where $\tilde g_{0t}$ is an unrestricted, possibly period-specific
production function and $\omega_t$ is latent productivity. Following
the proxy-variable approach
\citep{olley1996dynamics,levinsohn2003estimating,ackerberg2015identification},
assume that investment is strictly monotone in productivity conditional
on capital and labor, so that
\begin{align}
\omega_1
=
\tilde h_0(K_1,L_1,I_1)
\label{eq:prod_proxy}
\end{align}
for some unknown function $\tilde h_0$.

Productivity evolves according to
\begin{align}
\omega_2
=
\theta_0\omega_1+\xi_2.
\label{eq:prod_ar1_struct}
\end{align}
Let $\mathcal F_1$ denote the firm's information set after the
period-2 production inputs $(K_2,L_2)$ have been chosen but before the
period-2 productivity innovation $\xi_2$ and transitory output shock
$U_2$ are realized. Assume that
\[
(K_1,L_1,I_1,K_2,L_2)\subseteq\mathcal F_1
\]
and
\begin{align}
\mathbb E[U_1\mid K_1,L_1,I_1]&=0,
\label{eq:prod_timing1}
\\
\mathbb E[\xi_2+U_2\mid\mathcal F_1]&=0.
\label{eq:prod_timing2}
\end{align}
Thus, $I_1$ may respond to period-1 productivity, whereas the
period-2 inputs used in the instrument construction are predetermined
with respect to the innovation $\xi_2$ and the transitory shock $U_2$.
This timing structure is appropriate when these inputs are committed
before the period-2 innovation is observed; it is stronger than
production-function models in which flexible inputs respond
contemporaneously to productivity.

\paragraph{Reduced-form productivity.}
To map the problem into our general framework, define
\[
X_{11}\equiv(K_1,L_1),
\qquad
X_{12}\equiv(K_2,L_2),
\qquad
X_{21}\equiv(K_1,L_1,I_1).
\]
The first reduced-form regression projects output on the period-1 inputs
\[
g_{01}(X_{11})
\equiv
\mathbb E[Y_1\mid X_{11}],
\]
which by \eqref{eq:prod_struct} and \eqref{eq:prod_timing1} decomposes as
\[
g_{01}(X_{11})
=
\tilde g_{01}(K_1,L_1)
+
\mathbb E[\omega_1\mid K_1,L_1].
\]
Define the residualized generated outcome
\[
\tilde Y_1
\equiv
Y_1-g_{01}(X_{11})
\]
and its conditional mean
\begin{align}
h_0(X_{21})
&\equiv
\mathbb E[\tilde Y_1\mid X_{21}]
\nonumber\\
&=
\omega_1-\mathbb E[\omega_1\mid K_1,L_1].
\label{eq:prod_h0}
\end{align}
Hence $h_0$ recovers the component of period-1 productivity that is
not predictable from period-1 production inputs.

\paragraph{Constructed instruments.}
The remaining challenge is to construct an instrument that is
correlated with $h_0$ but orthogonal to the unrestricted production
function and to the predictable component of period-1 productivity.
Let
\[
\mathcal H
\equiv
\overline{
\left\{
a(X_{11})+b(X_{12}):
a\in L_2(X_{11}),\
b\in L_2(X_{12})
\right\}
}^{\,L_2}
\]
denote the closed additive space generated by period-1 and period-2
production inputs.

Let
\[
R\equiv r(X_{21})\in L_2
\]
be an observable period-1 productivity signal, where the transformation
$r$ is chosen by the researcher. Define
\begin{align}
q_0(X_{11},X_{12})
&\equiv
\Pi_{\mathcal H}R,
\label{eq:prod_q}
\\
Z_0
&\equiv
R-q_0(X_{11},X_{12}).
\label{eq:prod_Z}
\end{align}
Thus, $q_0$ is the $L_2$ projection of $R$ onto $\mathcal H$ and
$Z_0$ is its orthogonal residual. By construction,
\begin{align}
\mathbb E[Z_0 a(X_{11})]&=0
\quad
\text{for all }a\in L_2(X_{11}),
\label{eq:prod_orth11}
\\
\mathbb E[Z_0 b(X_{12})]&=0
\quad
\text{for all }b\in L_2(X_{12}).
\label{eq:prod_orth12}
\end{align}

Using \eqref{eq:prod_h0} and the productivity law of motion,
\begin{align}
Y_2-\theta_0 h_0(X_{21})
&=
\tilde g_{02}(K_2,L_2)
+
\theta_0
\mathbb E[\omega_1\mid K_1,L_1]
+
\xi_2+U_2.
\label{eq:prod_reduced}
\end{align}
The first two terms on the right-hand side belong to $\mathcal H$.
The constructed instrument $Z_0$ therefore eliminates both without
requiring either $\tilde g_{02}$ or
$\mathbb E[\omega_1\mid K_1,L_1]$ to be separately estimated.

\begin{proposition}[Constructed-instrument identification]
\label{prop:app}
Suppose equations
\eqref{eq:prod_struct}--\eqref{eq:prod_timing2} hold and let
$Z_0$ be defined by
\eqref{eq:prod_q}--\eqref{eq:prod_Z}. Then
\begin{align}
\mathbb E\left[
Z_0
\left\{
Y_2-\theta_0 h_0(X_{21})
\right\}
\right]
=0.
\label{eq:prod_moment}
\end{align}
If
\[
\mathbb E[Z_0h_0(X_{21})]\neq0,
\]
then $\theta_0$ is identified as
\[
\theta_0
=
\frac{\mathbb E[Z_0Y_2]}
{\mathbb E[Z_0h_0(X_{21})]}.
\]

Moreover, consider the class of constructed instruments
\[
\mathcal Z
\equiv
\left\{
r(X_{21})-\Pi_{\mathcal H}r(X_{21}):
r\in L_2(X_{21})
\right\}.
\]
There exists a relevant instrument in $\mathcal Z$ if and only if
\begin{align}
h_0(X_{21})\notin\mathcal H.
\label{eq:prod_relevance}
\end{align}
In particular, when \eqref{eq:prod_relevance} holds,
\[
Z_0^*
\equiv
h_0(X_{21})-\Pi_{\mathcal H}h_0(X_{21})
\]
satisfies
\[
\mathbb E[Z_0^*h_0(X_{21})]
=
\mathbb E[(Z_0^*)^2]
>0.
\]
\end{proposition}

Condition~\eqref{eq:prod_relevance} separates instrument validity from
relevance. Validity follows from the timing restrictions and the
orthogonalization in \eqref{eq:prod_Z}. Relevance requires the
investment proxy to contain variation in residualized period-1
productivity that cannot be reconstructed by adding an arbitrary
function of period-1 production inputs to an arbitrary function of
period-2 production inputs.

Importantly, proxy invertibility and instrument relevance are distinct
requirements. Strict monotonicity of investment implies that period-1
productivity can be recovered from $X_{21}$, but it does not imply
condition~\eqref{eq:prod_relevance}. If the residualized productivity
component $h_0(X_{21})$ can be reconstructed from an additive function
of $X_{11}$ and $X_{12}$, then every instrument obtained by
orthogonalizing a function of $X_{21}$ against $\mathcal H$ is
irrelevant. Hence identification requires the period-1 proxy to contain
productivity information that is not exhausted by current and future
production inputs.

The choice of the signal $R$ is important for relevance. Raw investment
need not itself generate a useful instrument. For example, under a
deterministic capital-accumulation equation, investment may belong to
the additive space $\mathcal H$, in which case its orthogonal residual
is identically zero. We therefore consider a dictionary
$\{r_j(X_{21})\}_{j=1}^J$ of nonlinear transformations and interactions
of $(K_1,L_1,I_1)$ and construct
\[
Z_{0j}
=
r_j(X_{21})
-
\Pi_{\mathcal H}r_j(X_{21}),
\qquad j=1,\ldots,J.
\]
For a fixed finite dictionary, the resulting vector
$Z_0=(Z_{01},\ldots,Z_{0J})'$ can be used directly in the
overidentified GMM framework developed above. For scalar $\theta_0$,
relevance requires
\[
\mathbb E[Z_0h_0(X_{21})]\neq0.
\]
At the population level, if the linear span of a sequence of candidate
signals is dense in $L_2(X_{21})$, condition~\eqref{eq:prod_relevance}
implies that some sufficiently rich finite subset contains relevant
instruments. In implementation, we use a fixed finite set of instruments chosen
ex ante from the candidate dictionary. Our asymptotic theory maintains strong identification. For a fixed
finite vector of constructed instruments, this corresponds here to
the nonsingularity condition in Assumption~\ref{ass:normality}(ii);
weak-instrument asymptotics are beyond the scope of the paper.
The projection on $\mathcal H$ must preserve its additive structure.
In particular, $q_0$ is the $L_2$ projection of $R$ onto
$L_2(X_{11})+L_2(X_{12})$, rather than the unrestricted conditional
expectation $\mathbb E[R\mid X_{11},X_{12}]$. This distinction is
important for relevance. Orthogonality to the additive space is
sufficient to remove the unrestricted period-2 production function and
the predictable component of period-1 productivity, while projecting
onto the larger space $L_2(X_{11},X_{12})$ may unnecessarily remove
additional variation in the period-1 proxy that is informative about
$h_0$.
\paragraph{Three-step representation.}
The construction fits our general framework with two first-step
regressions.

\bigskip \noindent
\textbf{First step.}
The reduced-form regression $g_{01}\in L_2(X_{11})$ is identified by
\[
\mathbb E\left[
\delta_{11}(X_{11})
\{Y_1-g_{01}(X_{11})\}
\right]
=0
\quad
\text{for all }\delta_{11}\in L_2(X_{11}).
\]
The instrument projection $q_0\in\mathcal H$ is identified by
\[
\mathbb E\left[
\delta_q(X_{11},X_{12})
\{R-q_0(X_{11},X_{12})\}
\right]
=0
\quad
\text{for all }\delta_q\in\mathcal H.
\]
Thus, $q_0$ is a first-step regression of the form considered in
Section~\ref{sec:set}, with the closed linear regression space
$\Delta_{1q}=\mathcal H$ rather than the unrestricted space
$L_2(X_{11},X_{12})$.
Consequently, using the product-space convention introduced in
Section~\ref{sec:set}, the first-step nuisance is
\[
g_0=(g_{01},q_0).
\]
The first component generates the residualized outcome
\[
\tilde Y_1(g_{01})
=
Y_1-g_{01}(X_{11}),
\]
while the second generates the instrument
\[
Z(q)=R-q(X_{11},X_{12}).
\]

\bigskip \noindent
\textbf{Second step.}
The second-step nuisance is
\[
h_0(X_{21})
=
\mathbb E[
\tilde Y_1(g_{01})\mid X_{21}
],
\]
or equivalently
\[
\mathbb E\left[
\delta_2(X_{21})
\left\{
\tilde Y_1(g_{01})-h_0(X_{21})
\right\}
\right]
=0
\quad
\text{for all }\delta_2\in L_2(X_{21}).
\]

\bigskip \noindent
\textbf{Third step.}
The identifying moment is
\[
m(W,g,h,\theta)
=
Z(q)
\left\{
Y_2-\theta h(X_{21})
\right\},
\]
so that at the truth Proposition~\ref{prop:app} gives
\[
\mathbb E[
m(W,g_0,h_0,\theta_0)
]
=0.
\]

This representation illustrates two distinct forms of
orthogonalization. The first is economic: $Z_0$ is constructed to be
orthogonal to arbitrary functions of period-1 and period-2 production
inputs. The second is statistical: the identifying moment must be made
locally robust to ML estimation of $g_{01}$, $q_0$, and $h_0$.

To derive the locally robust moment, define
\[
a_0(X_{21})
\equiv
\mathbb E[Z_0\mid X_{21}]
\]
and let
\[
\pi_0(X_{11},X_{12};\theta)
\equiv
\Pi_{\mathcal H}
\left[
Y_2-\theta h_0(X_{21})
\right].
\]
Both objects are well-defined functions of observables and nuisance
parameters. The function $a_0$ is the conditional mean of the
constructed instrument, while $\pi_0(\cdot;\theta)$ is the additive
projection of the third-step residual on $\mathcal H$.

\begin{proposition}\label{prop:prod_lr}
Consider the identifying moment in \eqref{eq:prod_moment}. A locally
robust moment for $\theta_0$ is
\begin{align}
\psi(W,g_0,h_0,\alpha_0,\theta)
&=
Z_0
\left\{
Y_2-\theta h_0(X_{21})
\right\}
\notag\\
&\quad
-\theta a_0(X_{21})
\left\{
Y_1-g_{01}(X_{11})-h_0(X_{21})
\right\}
\notag\\
&\quad
-\pi_0(X_{11},X_{12};\theta)
\left\{
R-q_0(X_{11},X_{12})
\right\}.
\label{eq:prod_lr}
\end{align}
The first-step Riesz representer is
$\alpha_{01}=(\alpha_{011},\alpha_{01q})$, where
\begin{align*}
\alpha_{011}(X_{11};\theta)&=0,
\\
\alpha_{01q}(X_{11},X_{12};\theta)
&=
-\pi_0(X_{11},X_{12};\theta),
\end{align*}
and the second-step Riesz representer is
\[
\alpha_{02}(X_{21};\theta)
=
-\theta a_0(X_{21}).
\]
\end{proposition}

The zero representer $\alpha_{011}$ has an economic interpretation.
Estimation of $g_{01}$ affects the downstream nuisance $h_0$, but the
resulting first-order effect on the identifying moment is proportional
to
\[
\mathbb E[Z_0\delta_{11}(X_{11})],
\]
which is zero by construction of the instrument. Thus, orthogonalizing
the instrument against $L_2(X_{11})$ also eliminates the first-order
effect of estimating the regression used to construct residualized
productivity.

The two nonzero corrections have distinct roles.
The term involving $a_0$ corrects for estimation of the downstream
productivity nuisance $h_0$, whereas the term involving $\pi_0$
corrects for estimation of the projection $q_0$ used to construct the
instrument. Hence the application combines a constructed instrument
with locally robust estimation of that instrument and of latent
productivity.

The procedure is directly implementable with cross-fitting. On each
training sample, estimate $g_{01}$ by regressing $Y_1$ on $X_{11}$,
estimate $q_0$ by an additive regression of $R$ on $X_{11}$ and
$X_{12}$, thereby enforcing the projection space $\mathcal H$, and form $\hat Z=R-\hat q$. Next, estimate $h_0$
by regressing $Y_1-\hat g_{01}(X_{11})$ on $X_{21}$ and estimate
$a_0(X_{21})=\mathbb E[Z_0\mid X_{21}]$. Finally, estimate the additive
projection $\pi_0(\cdot;\theta)$. Since
\[
\pi_0(X_{11},X_{12};\theta)
=
\Pi_{\mathcal H}Y_2
-
\theta
\Pi_{\mathcal H}h_0(X_{21}),
\]
the two projections on the right-hand side can be estimated once and
combined for each value of $\theta$. The resulting cross-fitted moment
in \eqref{eq:prod_lr} can then be used in the GMM estimator of
Section~\ref{sec:estim}.

The residualized outcome
\[
\tilde Y_1
=
Y_1-g_{01}(X_{11})
\]
is an instance of the generated-outcome map
$\varphi(W,g_0)=Y-g_0(X_1)$ of Remark~\ref{rem:resid}, here applied to latent productivity rather than to income or consumption. Two features distinguish the present application. First, the second-step correction no longer vanishes: unlike in the partially linear case, where Robinson's moment is already orthogonal to $h_0$, the moment in \eqref{eq:prod_moment} depends on $h_0$ through the identifying equation, so that $\alpha_{02}(X_{21};\theta)=-\theta a_0(X_{21})\neq0$. Second, the identifying instrument is itself generated by a first-step projection, and locally robust inference must therefore correct for two distinct first steps.
\subsection{A Life-Cycle Correction to Inequality of Opportunity}\label{sec:iop}
Inequality of opportunity (IOp) measures the extent to which socioeconomic outcomes depend on circumstances beyond individual control, such as sex, place of birth, or parental background \citep{roemer1998equality, ferreira2016equality}. Under equality of opportunity, outcomes would be independent of circumstances, so IOp can be assessed by how well circumstances predict outcomes. In practice, the literature considers types, which are groups of individuals who share the same circumstances. The ex-ante approach, which has dominated empirical applications \citep{brunori2026inherited}, measures the opportunities of each type by its average outcome, so that IOp is the inequality between these type averages \citep{ferreira2011measurement}. This inequality is usually reported relative to total inequality, so that IOp becomes the share of outcome inequality accounted for by circumstances. \par
IOp concerns lifetime economic status, yet available data rarely follow individuals over their working lives. Most studies observe a single income realization per person, or at best a few years, for individuals of different ages. Because income varies systematically over the life cycle, measured IOp is then distorted by the ages at which individuals happen to be observed. Following the intergenerational mobility literature \citep{jenkins1987snapshots, haider2006life}, we refer to this distortion as life-cycle bias.\par
How life-cycle bias can be addressed depends on the income data available. When long panels are available, averaging income over many years mitigates the bias \citep{aaberge2011measuring, bjorklund2012equality}, although it does not eliminate it. Such data are scarce, however, particularly in developing countries, where IOp is often of greatest concern \citep{brunori2025inequality}. When only one or a few income observations per person are available, a simple and appealing alternative is to purge income of its life-cycle component \citep{brunori2025inequality, brunori2026inherited}. This is typically done by regressing log income on a quadratic in age and using the residual plus a constant, or its exponential, as the outcome. Such a correction attempts to make individuals observed at different ages comparable without requiring repeated income observations. \par
The aforementioned age adjustment rests on two assumptions about the age profile, both of which can be relaxed. First, the profile is assumed to be quadratic. A low-dimensional fit is determined mainly by the ages at which most observations lie and can fit the profile poorly at the boundaries of the age range. Second, and more importantly, the age-profile is implicitly assumed to be common to all types. Earnings growth is steeper for individuals from more advantaged backgrounds \citep{nybom2016heterogeneous, halvorsen2022earnings}, so the shape of the age profile itself depends on circumstances. Capturing these interactions with many circumstances is a natural task for machine learning, and the resulting purged income becomes a generated outcome that falls within our framework.\par
A correction for life-cycle bias must account for two distinct ways in which age distorts the average income of a type. To make them explicit, let $Y$ denote log equivalized income, $A$ age, and $C$ circumstances, and let $g_0(a,c)=\mathbb{E}[Y\mid A=a,C=c]$ be the age--income profile of type $c$. The expected income of type $c$ is then
\begin{equation}\label{eq:iop_lc}
\mathbb{E}[Y\mid C=c]=\sum_{a} \underbrace{P(A=a\mid C=c)}_{\text{composition}}\;\underbrace{g_0(a,c)}_{\text{shape}},
\end{equation}
a weighted average of the type's profile, with weights given by its age distribution. Equation \eqref{eq:iop_lc} makes the two sources of life-cycle bias explicit. The first is composition: types may differ in their age distributions, so that types with identical profiles appear unequal. The second is shape: types may differ in their age--income profiles, so that the measured gap between them depends on the ages at which they are observed.\par
To address the two components of \eqref{eq:iop_lc}, we start with shape. Because profiles $g_0(\cdot,c)$ differ across types, the gap between two types depends on the ages at which it is measured. A meaningful comparison therefore evaluates the profile of every type at the same ages. Doing so requires two choices: which ages to use, and how to weight them. Let $\mathcal{R}$ be a set of reference ages with known weights $p_{\mathcal{R}}(b)\geq 0$, $b\in\mathcal{R}$, that sum to one. Evaluating the profile of type $c$ at these ages and averaging with these weights gives its reference income,
\begin{equation}\label{eq:iop_ref}
\bar g_{\mathcal{R}}(c)\coloneqq \sum_{b\in\mathcal{R}}p_{\mathcal{R}}(b)\,g_0(b,c),
\end{equation}
the expected income of type $c$ at the reference ages. Each type is evaluated along its own profile, but all types at the same stage of the life cycle. This component adapts to IOp the life-cycle estimator for the intergenerational elasticity of \citet{mello2025lifecycle}: to remove life-cycle bias in children's income, they predict it at common ages from profiles that vary with family background and use the predicted income as the outcome. We apply the same idea to inequality of opportunity, with circumstances in place of parental income.\par
The reference ages determine the stage of the life cycle at which IOp is measured. Reference ages spanning the whole working life compare types by their average income over that span, whereas a single age or mid-career ages compare them at that stage. Whatever the choice, identification requires every type to be observed at every reference age, so that the conditional mean $g_0(b,c)$ is well-defined. Root-$n$ inference requires, in addition, that no type be too rarely observed at any reference age, that is, $P(A=b\mid C)\geq\underline{\pi}>0$ for all $b\in\mathcal{R}$. In our application, the reference ages correspond to the working life observed in the data (25--59).\par
Consider next composition. In \eqref{eq:iop_lc}, the age distribution of a type matters only because the type's expected income varies with age. It therefore suffices to construct an outcome whose expected value within each type does not vary with age. This outcome must be defined at the individual level, since relative IOp also requires its total inequality. To construct it, write observed income as
\begin{equation*}
Y = g_0(A,C) + \varepsilon, \qquad \mathbb{E}[\varepsilon\mid A,C]=0,
\end{equation*}
where $g_0(A,C)$ is the expected income of the individual's type at their age and $\varepsilon$ is the individual's deviation from it. Of the two terms, only $g_0(A,C)$ varies with age in expectation, since $\varepsilon$ has mean zero at every age. Removing composition effects thus amounts to correcting $g_0(A,C)$ while leaving the individual's deviation $\varepsilon$ untouched.\par
Our proposed correction addresses both composition and shape distortions. To remove the effect of the former, we subtract $g_0(A,C)$ from observed income and keep $\varepsilon$. Subtracting it alone, however, would leave only $\varepsilon$, which has mean zero in every type, and would erase the differences between types we want to measure. We therefore add back a level for each type that does not depend on age, namely its reference income \eqref{eq:iop_ref}, which places every type at the same stage of the life cycle. The correction thus shifts each individual from the expected income of their type at their observed age to the expected income of the same type at the reference ages, while preserving their deviation from it. The result is the corrected (generated) income
\begin{equation}\label{eq:iop_purge}
\tilde Y = Y - g_0(A,C) + \bar g_{\mathcal{R}}(C) = \bar g_{\mathcal{R}}(C)+\varepsilon .
\end{equation}
Corrected income thus has two components: the reference income $\bar g_{\mathcal{R}}(C)$, which integrates the age profile of the individual's type over the reference ages and is therefore the same for every member of the type, whatever their observed age; and the individual's own deviation $\varepsilon$, which the correction leaves untouched and which has mean zero at every age. Hence $\mathbb{E}[\tilde Y\mid A=a,C=c]=\bar g_{\mathcal{R}}(c)$ for every age $a$, and applying the decomposition \eqref{eq:iop_lc} to $\tilde Y$ gives
\begin{equation*}
\mathbb{E}[\tilde Y\mid C=c]=\sum_{a} P(A=a\mid C=c)\,\bar g_{\mathcal{R}}(c)=\bar g_{\mathcal{R}}(c),
\end{equation*}
so a type's age distribution no longer affects its expected income. Subtracting $g_0(A,C)$ thus removes composition, and adding back $\bar g_{\mathcal{R}}(C)$ removes shape by evaluating every type at the same ages.  The correction standardizes the conditional mean of log income, not its dispersion: each individual keeps the deviation $\varepsilon$ of their observed age, so corrected income reflects the reference ages only insofar as the dispersion of $\varepsilon$ does not vary with age.\par A more comprehensive correction would also standardize this dispersion at the reference ages, evaluating $\mathbb{E}[\varepsilon^2\mid A,C]$ there for the variance of logs and $\mathbb{E}[e^{\varepsilon}\mid A,C]$ for the MLD. Both are regressions of generated outcomes on age and type, so our framework accommodates them. In our application, the dispersion of the deviations is nearly constant across ages, and the estimates with this additional standardization are close to ours and not significantly different at the 5\% level. We leave a full treatment of this extension for future research.\par
Equation \eqref{eq:iop_purge} nests the quadratic age adjustment used in the literature, which corresponds to imposing the two assumptions discussed above. Suppose first that the age profile is common to all types up to a level shift, $g_0(a,c)=\lambda(a)+f(c)$. Since the reference weights sum to one, the reference income becomes
\begin{equation*}
\bar g_{\mathcal{R}}(C)=\sum_{b\in\mathcal{R}}p_{\mathcal{R}}(b)\,\big[\lambda(b)+f(C)\big]=c_{\mathcal{R}}+f(C), \qquad c_{\mathcal{R}}=\sum_{b\in\mathcal{R}}p_{\mathcal{R}}(b)\,\lambda(b),
\end{equation*}
and substituting into \eqref{eq:iop_purge} gives
\begin{equation*}
\tilde Y = Y - \lambda(A) - f(C) + c_{\mathcal{R}} + f(C) = Y - \lambda(A) + c_{\mathcal{R}} .
\end{equation*}
The type-specific level $f(C)$ cancels and $c_{\mathcal{R}}$ is the same for every individual, so the correction reduces to removing a common age effect. If, in addition, the common profile is quadratic, $\lambda(a)=\beta_1 a+\beta_2 a^2$, with the intercept absorbed into $f(c)$, then
\begin{equation*}
\tilde Y = Y - \beta_1 A - \beta_2 A^2 + c_{\mathcal{R}} .
\end{equation*}
This is the form of the quadratic age adjustment used in the literature, up to the constant $c_{\mathcal{R}}$, which leaves relative IOp unchanged. In practice, $(\beta_1,\beta_2)$ are estimated by regressing $Y$ on a constant, age and age squared. Since $\mathbb{E}[Y\mid A]=\lambda(A)+\mathbb{E}[f(C)\mid A]$, this regression recovers $(\beta_1,\beta_2)$ when $\mathbb{E}[f(C)\mid A]$ does not vary with age, as when all types share the same age distribution. Otherwise, its slopes generally also absorb differences between types, and recovering the common profile requires conditioning on circumstances. Our correction thus keeps the logic of this adjustment and relaxes both of its assumptions. Relaxing them, however, turns the age profile into a high-dimensional object. With many circumstances, type-specific profiles must be estimated flexibly, so the corrected income $\tilde Y$ depends on an estimated and regularized first step. Errors in estimating $g_0$ therefore propagate both to the type averages and to the total inequality of corrected income, so treating $\tilde Y$ as observed ignores a first-order effect on relative IOp.\par
\paragraph{Three-step representation.}
The problem fits our framework as follows.

\bigskip \noindent
\textbf{First step.}
Let $X_1=(A,C)$ and define $g_0\in L_2(X_1)$ by
\[
\mathbb{E}[\delta_1(X_1)\{Y-g_0(X_1)\}]=0
\quad
\text{for all } \delta_1\in L_2(X_1),
\]
so that $g_0(A,C)=\mathbb{E}[Y\mid A,C]$ is the age--income profile of each type. The generated outcome is the corrected income \eqref{eq:iop_purge}, which evaluates the first step both at the individual's own age and at the reference ages,
\[
\tilde Y
=
\varphi(W,g_0)
=
Y-g_0(A,C)+\sum_{b\in\mathcal{R}}p_{\mathcal{R}}(b)\,g_0(b,C).
\]

\bigskip \noindent
\textbf{Second step.}
Let $X_2=C$ and $\Delta_2=L_2(X_2)$. Inequality indices are computed either on log income or on income levels, so let $S(g)=\tilde Y(g)$ or $S(g)=e^{\tilde Y(g)}$, respectively, denote corrected income on the scale of the index. The second step gives the type averages
\[
h_0(X_2)
=
\mathbb{E}[S(g_0)\mid X_2].
\]
On the log scale, $h_0(C)=\mathbb{E}[\tilde Y\mid C]$ equals the reference income $\bar g_{\mathcal{R}}(C)$ by construction.\par

\bigskip \noindent
\textbf{Third step.}
Relative IOp is the inequality between type averages as a share of total inequality, both measured with the same index. Most relative indices used in the literature are smooth functions of a few moments,
\[
I(V)=\Lambda\big(\mathbb{E}[\kappa(V)]\big),
\]
for a known vector of functions $\kappa(\cdot)$ and a smooth function $\Lambda$. Relative IOp is then
\[
\theta_0=\frac{\Lambda(\bar\kappa_{N0})}{\Lambda(\bar\kappa_{D0})},
\qquad
\bar\kappa_{N0}=\mathbb{E}\big[\kappa\big(h_0(C)\big)\big],
\quad
\bar\kappa_{D0}=\mathbb{E}\big[\kappa\big(S(g_0)\big)\big],
\]
where $(\bar\kappa_0,\theta_0)$, with $\bar\kappa_0=(\bar\kappa_{N0},\bar\kappa_{D0})$, is identified by $\mathbb{E}[m(W,g_0,h_0,\bar\kappa_0,\theta_0)]=0$ with
\[
m(W,g,h,\bar\kappa,\theta)=
\begin{pmatrix}
\kappa\big(h(C)\big)-\bar\kappa_N\\[2pt]
\kappa\big(S(g)\big)-\bar\kappa_D\\[2pt]
\Lambda(\bar\kappa_N)-\theta\,\Lambda(\bar\kappa_D)
\end{pmatrix}.
\] All indices below are invariant to proportional changes in income, equivalently to additive shifts in log income, so constants such as $c_{\mathcal{R}}$ are irrelevant.

\smallskip \noindent
\emph{Variance of logs.} With $S(g)=\tilde Y(g)$, $\kappa(v)=(v,v^2)$ and $\Lambda(a_1,a_2)=a_2-a_1^2$, $I(V)=\operatorname{Var}(V)$ and
\[
\theta_0=\frac{\operatorname{Var}\big(h_0(C)\big)}{\operatorname{Var}(\tilde Y)},
\]
the share of the variance of corrected log income accounted for by circumstances. Since $\mathbb{E}[h_0(C)]=\mathbb{E}[\tilde Y]=\mu_0$, relative IOp and the mean are identified by $\mathbb{E}[m(W,g_0,h_0,\theta_0,\mu_0)]=0$ with
\[
m(W,g,h,\theta,\mu)=
\begin{pmatrix}
\big(h(C)-\mu\big)^2-\theta\,\big(\tilde Y(g)-\mu\big)^2\\[2pt]
\tilde Y(g)-\mu
\end{pmatrix}.
\]

\smallskip \noindent
\emph{Mean log deviation.} With $S(g)=e^{\tilde Y(g)}$, $\kappa(v)=(v,\ln v)$ and $\Lambda(a_1,a_2)=\ln a_1-a_2$, $I(V)=\ln\mathbb{E}[V]-\mathbb{E}[\ln V]$. Corrected income in levels is $e^{\tilde Y}=e^{\,Y-g_0(A,C)+\bar g_{\mathcal{R}}(C)}$, the exponentiated outcome of \citet{brunori2025inequality}, with type averages $h_0(C)=\mathbb{E}[e^{\tilde Y}\mid C]$. Since $\mathbb{E}[h_0(C)]=\mathbb{E}[e^{\tilde Y}]=\mu_0$,
\[
\theta_0=\frac{\ln\mu_0-\mathbb{E}\big[\ln h_0(C)\big]}{\ln\mu_0-\mathbb{E}[\tilde Y]},
\]
and $(\theta_0,\mu_0)$ is identified by $\mathbb{E}[m(W,g_0,h_0,\theta_0,\mu_0)]=0$ with
\[
m(W,g,h,\theta,\mu)=
\begin{pmatrix}
\ln\mu-\ln h(C)-\theta\,\big(\ln\mu-\tilde Y(g)\big)\\[2pt]
e^{\tilde Y(g)}-\mu
\end{pmatrix}.
\]
\par
Under overlap, the three steps identify relative IOp: the profile $g_0$ is identified at every reference age because every type is observed there, the type averages $h_0$ are identified from corrected income, and $\theta_0$ is a smooth function of the identified means $\bar\kappa_0$. For every index in this class, the moment depends on the first step through two channels. The numerator depends on $g_0$ only through the type averages $h_0$, whereas the denominator depends on corrected income $S(g)$ directly. Relative IOp is therefore an instance of the general case in Remark~\ref{rem:linear}, in which a single first-step regression enters the moment both directly and through the second step.\par
The quadratic age adjustment fits the same setting, with the restricted first-step space $\Delta_1$ spanned by $\{1,A,A^2\}$. Because this first step is low-dimensional and estimated at the parametric rate, it introduces no regularization bias. Moreover, since the Riesz representer lies in $\Delta_1$, the normal equations make its correction to the point estimate zero, up to cross-fitting. In general, however, its estimation still contributes to the asymptotic variance, which the locally robust standard errors account for. The main limitation of the quadratic adjustment is therefore misspecification, which arises whenever the age profile is not quadratic or not common to all types, or when the age distribution differs across types. Flexible estimation by machine learning avoids this misspecification but introduces regularization bias. Local robustness removes this bias and preserves valid inference, with confidence intervals that account for first-step uncertainty.\par
Following Theorem~\ref{thm:lr}, the locally robust moment augments the identifying moment $m$ with the first- and second-step influence functions. Both depend on how corrected income responds to the first step. For a first-step perturbation $\delta_1\in L_2(A,C)$, corrected income changes by
\[
D_gS(g_0)[\delta_1]=\dot S\,\Big\{\sum_{b\in\mathcal{R}}p_{\mathcal{R}}(b)\,\delta_1(b,C)-\delta_1(A,C)\Big\},
\]
where $\dot S=1$ on the log scale and $\dot S=e^{\tilde Y}$ on income levels. Let $p_0(b\mid c)=P(A=b\mid C=c)$ denote the age distribution of type $c$, and define the weights
\[
w_{\mathcal{R}}(A,C)=\frac{\mathbf{1}\{A\in\mathcal{R}\}\,p_{\mathcal{R}}(A)}{p_0(A\mid C)},
\]
which reweight the age distribution of each type to the reference distribution. They are bounded under overlap and satisfy $\mathbb{E}[w_{\mathcal{R}}(A,C)\,\delta_1(A,C)\mid C]=\sum_{b\in\mathcal{R}}p_{\mathcal{R}}(b)\,\delta_1(b,C)$, so they express the evaluation of the first step at the reference ages as an evaluation at the observed ages; in particular, $\mathbb{E}[w_{\mathcal{R}}(A,C)\mid C]=1$. Finally, let $\kappa'(\cdot)$ denote the vector of derivatives of $\kappa(\cdot)$ and define
\[
Q_N(A,C)=\mathbb{E}\big[\kappa'\big(h_0(C)\big)\,\dot S\mid A,C\big],
\qquad
Q_D(A,C)=\mathbb{E}\big[\kappa'\big(S(g_0)\big)\,\dot S\mid A,C\big],
\]
which measure how the numerator and denominator moments respond to a marginal change in the corrected income of an individual of type $C$ observed at age $A$.
\begin{proposition}\label{prop:iop}
Suppose that Assumption~\ref{ass:lr} and the overlap condition hold, that $\kappa$ and $\Lambda$ are continuously differentiable with $\Lambda(\bar\kappa_{D0})>0$, and that $\kappa(S(g_0))$, $\kappa'(h_0(C))$, $\kappa'(S(g_0))\dot S$ and $\kappa'(h_0(C))\dot S$ have finite second moments. The locally robust moment of Theorem~\ref{thm:lr} for $(\bar\kappa_0,\theta_0)$ is
\[
\psi(W,g,h,\bar\kappa,\theta)=m(W,g,h,\bar\kappa,\theta)+\alpha_{02}(C)\big\{S(g)-h(C)\big\}+\alpha_{01}(A,C)\big\{Y-g(A,C)\big\},
\]
where, with one representer for each component of $m$,
\[
\alpha_{02}(C)=
\begin{pmatrix}
\kappa'\big(h_0(C)\big)\\[2pt]
0\\[2pt]
0
\end{pmatrix},
\qquad
\alpha_{01}(A,C)=
\begin{pmatrix}
w_{\mathcal{R}}(A,C)\,\mathbb{E}[Q_N\mid C]-Q_N(A,C)\\[2pt]
w_{\mathcal{R}}(A,C)\,\mathbb{E}[Q_D\mid C]-Q_D(A,C)\\[2pt]
0
\end{pmatrix}.
\]
\end{proposition}

The second-step correction $\alpha_{02}(C)\{S(g)-h(C)\}$ accounts for the machine learning estimation of the type averages. It is the analogue, for indices of the form $\Lambda(\mathbb{E}[\kappa(\cdot)])$, of the corrections that \citet{escanciano2025debiased} derive for the Gini, generalized entropy and Atkinson indices of predicted type averages, and coincides with theirs for the mean log deviation; here it is applied to corrected rather than observed income. Without a life-cycle correction, the first step disappears and $\psi$ reduces to this second-step debiasing. The first-step correction $\alpha_{01}$ is specific to the generated outcome. A first-step error enters corrected income twice: with a minus sign at the observed age, which contributes $-Q(A,C)$, and with a plus sign at the reference ages, through the reference income. Averaged within the type, the latter contributes $\mathbb{E}[Q\mid C]$, and expressed at the ages at which the type is observed it becomes $w_{\mathcal{R}}(A,C)\,\mathbb{E}[Q\mid C]$. Each component of $\alpha_{01}$ is the difference between the two: through $Q_N$ it captures the indirect effect of the first step through the type averages, and through $Q_D$ its direct effect on the total inequality of corrected income. For the variance of logs and the mean log deviation, the two indices used in our application, the representers have closed forms.

\begin{corollary}\label{cor:iop}
Under the conditions of Proposition~\ref{prop:iop} other than Assumption~\ref{ass:lr}, which holds for the variance of logs and, under the additional condition in (ii), for the mean log deviation, locally robust moments for $(\theta_0,\mu_0)$ are as follows, where the first component identifies $\theta_0$, the second identifies $\mu_0$, and the dependence of the representers on $(\theta,\mu)$ is suppressed.
\begin{enumerate}[(i)]
\item \emph{Variance of logs.}
\begin{align*}
\psi_1(W,g,h,\theta,\mu)&=\big(h(C)-\mu\big)^2-\theta\big(\tilde Y(g)-\mu\big)^2+2\big(h(C)-\mu\big)\big(\tilde Y(g)-h(C)\big)\\
&\quad+\alpha_{01,1}(A,C)\big\{Y-g(A,C)\big\},\\
\psi_2(W,g,\mu)&=\tilde Y(g)-\mu+\alpha_{01,2}(A,C)\big\{Y-g(A,C)\big\},
\end{align*}
with
\[
\alpha_{01,1}(A,C)=2(1-\theta)\big(h_0(C)-\mu\big)\big(w_{\mathcal{R}}(A,C)-1\big),
\qquad
\alpha_{01,2}(A,C)=w_{\mathcal{R}}(A,C)-1.
\]
\item \emph{Mean log deviation.} Suppose, in addition, that $(A,C)$ has finite support and $h_0(C)\geq c>0$ almost surely, and let $H_0(A,C)=\mathbb{E}[e^{\tilde Y}\mid A,C]$, so that $\mathbb{E}[H_0\mid C]=h_0(C)$. Then
\begin{align*}
\psi_1(W,g,h,\theta,\mu)&=\ln\mu-\ln h(C)-\theta\big(\ln\mu-\tilde Y(g)\big)-\frac{e^{\tilde Y(g)}-h(C)}{h(C)}\\
&\quad+\alpha_{01,1}(A,C)\big\{Y-g(A,C)\big\},\\
\psi_2(W,g,\mu)&=e^{\tilde Y(g)}-\mu+\alpha_{01,2}(A,C)\big\{Y-g(A,C)\big\},
\end{align*}
with
\begin{align*}
\alpha_{01,1}(A,C)&=\frac{H_0(A,C)}{h_0(C)}-w_{\mathcal{R}}(A,C)+\theta\big(w_{\mathcal{R}}(A,C)-1\big),\\
\alpha_{01,2}(A,C)&=w_{\mathcal{R}}(A,C)\,h_0(C)-H_0(A,C).
\end{align*}
\end{enumerate}
\end{corollary}
\par The first-step representers also have a simple interpretation. Since $\mathbb{E}[w_{\mathcal{R}}\mid C]=1$, every component of $\alpha_{01}$ has mean zero within each type, so a first-step error that shifts the whole profile of a type up or down leaves the moment unchanged: it is subtracted at the observed age and added back at the reference ages. The correction responds only to errors in the shape of the profiles, which is precisely what machine learning must estimate flexibly. For the variance of logs, $\alpha_{01,1}$ weights each individual by how far their type sits from the mean, $h_0(C)-\mu_0$, and by how far the type's observed age composition departs from the reference stage of the life cycle, $w_{\mathcal{R}}-1$. A type observed exactly at the reference age distribution has $w_{\mathcal{R}}\equiv1$ and contributes nothing, because its subtraction and its add-back are evaluated at the same ages. Age thus plays the role of a treatment set to counterfactual values: $w_{\mathcal{R}}$ is the corresponding inverse-probability weight, and overlap plays the role of positivity.

\begin{remark}[Other inequality indices]\label{rem:iop_indices}
The same structure covers the generalized entropy and Atkinson families on income levels, for example GE(2), with $\kappa(v)=(v,v^2)$ and $\Lambda(a_1,a_2)=(a_2/a_1^2-1)/2$, or the Atkinson index with parameter $\epsilon$, with $\kappa(v)=(v,v^{1-\epsilon})$ and $\Lambda(a_1,a_2)=1-a_2^{1/(1-\epsilon)}/a_1$; only the auxiliary regressions entering the correction change, and Assumption~\ref{ass:lr} holds under the finite support of Corollary~\ref{cor:iop}(ii). For these families, the second-step correction follows from \citet{escanciano2025debiased}, so that only the first-step correction is new. The Gini coefficient, which depends on pairs of observations, can be accommodated by combining our correction with the debiased U-statistic theory of \cite{escanciano2025debiased}.
\end{remark}
\begin{remark}[Inference]\label{rem:iop_inference}
The conditions of Theorem~\ref{thm:normality} can be verified along the lines of Corollary~\ref{cor:plm_normality} for the partially linear model. Beyond finite fourth moments, the main requirements are that the first step $\hat g$, the type averages $\hat h$, the estimated age distribution $\hat p$ and the auxiliary regressions entering $\hat\alpha_{01}$ be consistent in $L_4$, that $\sqrt{n}\,(\|\hat g-g_0\|_2^2+\|\hat h-h_0\|_2^2)=o_p(1)$, and that $\sqrt{n}\,\|\hat\alpha_{01}-\alpha_{01}\|_2\,\|\hat g-g_0\|_2=o_p(1)$. Overlap keeps the estimated weights bounded, and since $\alpha_{02}$ is a smooth function of $h_0$, the corresponding product condition for the second step follows from the rate on $\hat h$. Inference with the MLD requires, in addition, the finite support in Corollary~\ref{cor:iop}(ii), that $\hat h\geq c$ with probability approaching one, and that $e^{\tilde Y}$ has finite fourth moments. Finite support bounds $H_0$ and the sup norm of the first-step error by its $L_2$ norm, so the remainder of the MLD moment is of order $\|\hat g-g_0\|_2^2$. 
\end{remark}
In the general formulation of Proposition~\ref{prop:iop}, the last component of $\psi$ is degenerate, so $\Psi$ is singular, which Theorem~\ref{thm:normality} allows.
\section{Estimation, Inference, and Finite-Sample Performance}\label{sec:estim}
\subsection{Estimation and Inference}
The orthogonal moment $\psi$ in Theorem~\ref{thm:lr} requires consistent estimators of the nuisance parameters $g_0$, $h_0$, and the Riesz representers $\alpha_0\equiv(\alpha_{01},\alpha_{02})$. We suppress the dependence of $\alpha_0(\theta)$ and
$\hat{\alpha}_\ell(\theta)$ on $\theta$ whenever no confusion arises. In our examples, the orthogonality conditions \eqref{eq:riesz1}--\eqref{eq:riesz2} admit analytical solutions, so consistent plug-in estimators are available directly from consistent estimators of the nuisance parameters they depend on. In settings where analytical expressions for the Riesz representers are unavailable or cumbersome to derive, automatic estimation \citep{chernozhukov2022locally} can be used in their place, without any modification to the locally robust moment construction of Theorem~\ref{thm:lr}.

To avoid own-observation bias and circumvent the need to verify Donsker-type conditions, we use cross-fitting: the sample
$(W_i)_{i=1}^n$ is partitioned into $L$ folds
$\mathcal I_\ell$, $\ell=1,\ldots,L$, and for each fold we obtain
estimators $\hat g_\ell$, $\hat h_\ell$, and
$\hat\alpha_\ell\equiv(\hat\alpha_{1\ell},\hat\alpha_{2\ell})$
using only observations outside $\mathcal I_\ell$.

Let
\[
\hat S_{i\ell}\equiv S(\hat g_\ell)(W_i)
\]
denote the cross-fitted generated outcome evaluated at observation
$i$. The debiased sample moment function is then
\begin{align}
\label{eq:sample_moment}
\hat\psi(\theta)
\equiv
\frac{1}{n}
\sum_{\ell=1}^{L}
\sum_{i\in\mathcal I_\ell}
\hat\psi_{i\ell}(\theta),
\end{align}
where
\begin{align*}
\hat\psi_{i\ell}(\theta)
\equiv
m(W_i,\hat g_\ell,\hat h_\ell,\theta)
+
\hat\alpha_{1\ell}(X_{1i})'\bigl(Y_i-\hat g_\ell(X_{1i})\bigr)
+
\hat\alpha_{2\ell}(X_{2i})'\bigl(\hat S_{i\ell}-\hat h_\ell(X_{2i})\bigr).
\end{align*}
When there is more than one moment condition, each component of $m$ is debiased by its own Riesz representers. The debiased GMM estimator is
\begin{align}
\label{eq:gmm}
\hat{\theta} = \argmin_{\theta \in \Theta}\,  \hat{\psi}(\theta)'\hat{\Upsilon}\,\hat{\psi}(\theta),
\end{align}
where $\hat{\Upsilon}$ is a positive semi-definite weighting matrix of dimension $q \times q$. \par
\bigskip
\textbf{Example: PLM with generated dependent variable, cont.}
In the partially linear model, the moment condition is exactly
identified and linear in $\theta_0$, so the locally robust moment in
equation~\eqref{eq:lr_plm_general_main} gives $\theta_0$ in closed
form,
\begin{align*}
\theta_0
&=
B^{-1}C,
\nonumber
\\
B
&\coloneqq
\mathbb{E}\left[
(X_3-h_{0X}(X_2))
(X_3-h_{0X}(X_2))'
\right],
\nonumber
\\
C
&\coloneqq
\mathbb E\left[
(\tilde Y-h_{0Y}(X_2))
(X_3-h_{0X}(X_2))
+
\alpha_{01}(X_1)(Y-\tilde Y)
\right],
\end{align*}
and the debiased GMM estimator in equation~\eqref{eq:gmm} reduces to
the sample analogue
\begin{align}
\label{eq:plm_estimator}
\hat\theta
=
\hat B^{-1}\hat C,
\end{align}
where
\begin{align*}
\hat B
&\coloneqq
\frac{1}{n}
\sum_{\ell=1}^L
\sum_{i\in\mathcal I_\ell}
\bigl(
X_{3i}-\hat h_{X\ell}(X_{2i})
\bigr)
\bigl(
X_{3i}-\hat h_{X\ell}(X_{2i})
\bigr)',
\\
\hat C
&\coloneqq
\frac{1}{n}
\sum_{\ell=1}^L
\sum_{i\in\mathcal I_\ell}
\Big[
\bigl(
\hat g_\ell(X_{1i})
-
\hat h_{Y\ell}(X_{2i})
\bigr)
\bigl(
X_{3i}-\hat h_{X\ell}(X_{2i})
\bigr)
\\
&\hspace{4cm}
+
\hat\alpha_{01\ell}(X_{1i})
\bigl(
Y_i-\hat g_\ell(X_{1i})
\bigr)
\Big].
\end{align*}
Here
\[
\hat h_{Y\ell}(X_2)
=
\hat{\mathbb E}_\ell
[
\hat g_\ell(X_1)\mid X_2
],
\qquad
\hat h_{X\ell}(X_2)
=
\hat{\mathbb E}_\ell
[
X_3\mid X_2
],
\]
where $\hat{\mathbb E}_\ell$ denotes ML estimators of the indicated
conditional means. The estimator $\hat\alpha_{01\ell}$ targets
\[
\alpha_{01}
=
\Pi_{\Delta_1}\{X_3-h_{0X}(X_2)\},
\]
with each fitted component belonging to $\Delta_1$.
All nuisance functions are estimated using observations outside
fold $\mathcal I_\ell$ and satisfy the convergence conditions
specified below.

We now establish consistency and asymptotic normality of $\hat{\theta}$ in the general case of (\ref{eq:gmm}). Both results follow the proof strategy of \cite{chernozhukov2022locally}, Theorem A3 for consistency and Theorem 9 for asymptotic normality, adapted to the compositional setting in which $\hat h_\ell$ is constructed from $S(\hat g_\ell)$. This dependence is handled through joint nuisance and remainder conditions rather than by treating the two nuisance estimators as independent first steps.\par
Consistency follows from standard M-estimation theory adapted to the locally robust framework. The following regularity conditions are required.
\begin{assumption}
\label{ass:consistency}\leavevmode
\begin{enumerate}[(i)]
\item The parameter space $\Theta \subset \mathbb{R}^p$ is compact and  $\theta_0$ is the unique solution to the moment condition $\mathbb{E}[\psi(W, g_0, h_0, \alpha_0, \theta)] = 0$ over $\theta \in \Theta$.
\item $\mathbb{E}[\|\psi(W, g_0, h_0, \alpha_0, \theta_0)\|] < \infty$ and $\int \|m(w, \hat{g}_\ell, \hat{h}_\ell, \theta) - m(w, g_0, h_0, \theta)\|  dF_0(w) \overset{p}{\to} 0$ for all $\theta \in \Theta$.
\item There exist $C>0$ and $d(W)$ with
$\mathbb{E}[d(W)]<\infty$ such that, almost surely,
for all $\theta,\tilde{\theta}\in\Theta$,
\[
\|\psi(W,g_0,h_0,\alpha_0,\theta)
-\psi(W,g_0,h_0,\alpha_0,\tilde{\theta})\|
\leq d(W)\|\theta-\tilde{\theta}\|^{1/C}.
\]
Moreover, for each fold $\ell$,
\[
\int\sup_{\theta\in\Theta}
\|\psi(w,\hat{g}_\ell,\hat{h}_\ell,\hat{\alpha}_\ell,\theta)
-\psi(w,g_0,h_0,\alpha_0,\theta)\|\,dF_0(w)
\overset{p}{\to}0.
\]
\item For each fold $\ell$, the estimated Riesz representers satisfy \[ \hat{\alpha}_{1\ell}\in\Delta_1^q, \qquad \hat{\alpha}_{2\ell}\in\Delta_2^q \] with probability approaching one. Moreover, with $\phi=\phi_1+\phi_2$,
\begin{align*}
\int \|\phi(w, \hat{g}_\ell, h_0, \alpha_0, \theta_0) -  \phi(w, g_0, h_0, \alpha_0, \theta_0)\|^2 dF_0(w) &\overset{p}{\to} 0,\\
\int \|\phi(w, g_0, \hat{h}_\ell, \alpha_0, \theta_0) -  \phi(w, g_0, h_0, \alpha_0, \theta_0)\|^2 dF_0(w) &\overset{p}{\to} 0,\\
\int \|\phi(w, g_0, h_0, \hat{\alpha}_\ell, \theta_0) - \phi(w, g_0, h_0, \alpha_0, \theta_0)\|^2 dF_0(w) &\overset{p}{\to} 0.
\end{align*}
\item For each fold $\ell$,
$\frac{1}{\sqrt{n}}\sum_{i\in\mathcal{I}_\ell}
\hat{\Delta}_\ell(W_i)\overset{p}{\to}0$, where
\begin{align*}
\hat{\Delta}_\ell(w)\coloneqq{}&
\psi(w,\hat{g}_\ell,\hat{h}_\ell,\hat{\alpha}_\ell,\theta_0)
-\psi(w,\hat{g}_\ell,\hat{h}_\ell,\alpha_0,\theta_0)\\
&-\psi(w,g_0,h_0,\hat{\alpha}_\ell,\theta_0)
+\psi(w,g_0,h_0,\alpha_0,\theta_0).
\end{align*}
\item $\hat{\Upsilon} \overset{p}{\to} \Upsilon$ positive definite.
\end{enumerate}
\end{assumption}
The identification condition in $(i)$ is standard. The $L_1$ continuity condition in $(ii)$ ensures the difference between the moment evaluated at estimated and true nuisance parameters becomes negligible as $n \to \infty$. Condition~$(iii)$ ensures continuity of the population moment
and uniform approximation of the estimated score by its
true-nuisance counterpart, yielding the uniform convergence
needed for consistency. Conditions~$(iv)$ and $(v)$ are standard mean-square
consistency and remainder conditions, and condition~$(vi)$
imposes convergence of the weighting matrix. The dependence of
$\hat h_\ell$ on $\hat g_\ell$ through the generated outcome is
handled by these joint nuisance and remainder conditions rather
than by imposing a separate product-rate restriction at the
consistency stage.
\begin{lemma}[Consistency]
\label{lem:consistency}
Under Assumptions~\ref{ass:lr} and \ref{ass:consistency}, the debiased GMM estimator $\hat{\theta}$ in equation \eqref{eq:gmm} satisfies
\[
\hat{\theta} \overset{p}{\to} \theta_0.
\]
\end{lemma}
Having established consistency, asymptotic normality requires additional structure on the moment integrability and Jacobian stability beyond Assumption~\ref{ass:consistency}. Let $\bar{\psi}$ denote the expected debiased moment evaluated at estimated nuisances, $\bar{\psi}(\hat{g}_\ell, \hat{h}_\ell, \alpha_0, \theta_0) \equiv \mathbb{E}[\psi(W, \hat{g}_\ell, \hat{h}_\ell, \alpha_0, \theta_0)]$.
\begin{assumption}
\label{ass:normality}\leavevmode
\begin{enumerate}[(i)]
\item $\mathbb{E}[\|\psi(W, g_0, h_0, \alpha_0, \theta_0)\|^2] < \infty$. Let $\Psi\equiv\mathbb{E}[\psi(W,g_0,h_0,\alpha_0,\theta_0)\psi(W,g_0,h_0,\alpha_0,\theta_0)']$, which may be singular.
\item $\theta_0\in\operatorname{int}(\Theta)$. The Jacobian
\[
G\coloneqq
\mathbb{E}\left[
\partial_\theta
\psi(W,g_0,h_0,\alpha_0,\theta_0)
\right]
\]
exists and $G'\Upsilon G$ is nonsingular.

\item Let $\eta=(g,h,\alpha)$ and
$\eta_0=(g_0,h_0,\alpha_0)$. For all
$\|\eta-\eta_0\|$ sufficiently small,
$\psi(W,\eta,\theta)$ is differentiable with respect to $\theta$
with probability approaching one, and there exist $C>0$ and
$d(W)$ such that, for all $\theta\in\Theta$,
\[
\left\|
\frac{\partial\psi(W,\eta,\theta)}{\partial\theta}
-
\frac{\partial\psi(W,\eta,\theta_0)}{\partial\theta}
\right\|
\leq
d(W)\|\theta-\theta_0\|^{1/C},
\qquad
\mathbb{E}[d(W)]<C.
\]

\item For each fold $\ell$,
\[
\int
\left\|
\frac{\partial
\psi(w,\hat g_\ell,\hat h_\ell,\hat\alpha_\ell,\theta_0)}
{\partial\theta}
-
\frac{\partial
\psi(w,g_0,h_0,\alpha_0,\theta_0)}
{\partial\theta}
\right\|
dF_0(w)
\overset{p}{\to}0.
\]
\item For each fold $\ell$,
\begin{align*}
\int \|m(w,\hat{g}_\ell,\hat{h}_\ell,\theta_0)
-m(w,g_0,h_0,\theta_0)\|^2\,dF_0(w)
&\overset{p}{\to}0,\\
\int \|\hat{\Delta}_\ell(w)\|^2\,dF_0(w)
&\overset{p}{\to}0.
\end{align*}
\item For each fold $\ell$, one of the following holds: either $(a)$ there exists $C > 0$ such that, with probability tending to one,
\[
\left|\bar{\psi}(\hat{g}_\ell, \hat{h}_\ell, \alpha_0, \theta_0)\right| \leq C\left(\|\hat{g}_\ell - g_0\|_2^2 + \|\hat{h}_\ell - h_0\|_2^2\right),\]
and $\sqrt{n}\left(\|\hat{g}_\ell - g_0\|_2^2 + \|\hat{h}_\ell - h_0\|_2^2\right) = o_p(1)$; or $(b)$
\[\sqrt{n}\,\bar{\psi}(\hat{g}_\ell, \hat{h}_\ell, \alpha_0, \theta_0) \overset{p}{\to} 0.\]
\end{enumerate}
\end{assumption}
Condition~$(i)$ strengthens the integrability requirement in Assumption~\ref{ass:consistency}$(ii)$ from $L_1$ to $L_2$, needed to apply a central limit theorem, but allows $\Psi$ to be singular, as in the general formulation of Section~\ref{sec:iop}. Condition~$(ii)$ guarantees the GMM problem is well-posed; since the sandwich does not invert $\Psi$, inference on $c'\theta_0$ only requires $c'Vc>0$. Since the influence adjustments have mean zero for every admissible $\alpha_0(\theta)$, 
\[ G = \mathbb{E}\left[ \partial_\theta \psi(W,g_0,h_0,\alpha_0,\theta_0) \right] = \mathbb{E}\left[ \partial_\theta m(W,g_0,h_0,\theta_0) \right]. \] Conditions~$(iii)$ and~$(iv)$ control local stability of the Jacobian with respect to $\theta$ and the nuisance parameters respectively, ensuring that first-stage estimation error does not distort the asymptotic behavior of $\hat{\theta}$. Condition~$(v)$ supplies mean-square control of the identifying
moment and the interaction remainder, complementing Assumption~\ref{ass:consistency}$(iv)$ on the influence corrections. Condition~$(vi)$ corresponds to Assumption~6.8 of \cite{escanciano2023automatic} and controls the first-order bias of the debiased moment when nuisance functions are estimated jointly: it requires either a quadratic remainder bound on the expected moment at estimated  nuisances, or directly that $\sqrt{n}\bar{\psi}(\hat{g}_\ell, \hat{h}_\ell) \overset{p}{\to} 0$. Either condition ensures the $\sqrt{n}$-normalized expected remainder vanishes, which is not needed for Lemma~\ref{lem:consistency} but is required for asymptotic normality. Hadamard differentiability alone does not imply a quadratic remainder bound. For the PLM, an exact expansion yields such a bound, and condition~$(a)$ follows from
Assumption~\ref{ass:plm}$(iv)$; see the proof of
Corollary~\ref{cor:plm_normality}.
\begin{theorem}[Asymptotic Normality]\label{thm:normality}
Under Assumptions~\ref{ass:lr}--\ref{ass:normality}, the debiased GMM estimator $\hat{\theta}$ in equation \eqref{eq:gmm} satisfies
\[\sqrt{n}(\hat{\theta} - \theta_0) \xrightarrow{d} \mathcal{N}(0, V),
\] where $V = (G'\Upsilon G)^{-1}(G'\Upsilon\Psi\Upsilon G)(G'\Upsilon G)^{-1}$, $\Psi = \mathbb{E}[\psi(W, g_0, h_0, \alpha_0, \theta_0)\psi(W, g_0, h_0, \alpha_0, \theta_0)']$, and $\Upsilon$ is the probability limit of $\hat{\Upsilon}$. Moreover, $\hat{V} \overset{p}{\to} V$, and $\sqrt{n}\,c'(\hat{\theta}-\theta_0)/(c'\hat{V}c)^{1/2}\xrightarrow{d}\mathcal{N}(0,1)$ for every fixed $c$ with $c'Vc>0$.
\end{theorem}
For inference, let
\[\hat G\equiv\frac{\partial\hat\psi(\hat\theta)}{\partial\theta},\qquad\hat\Psi
\equiv\frac{1}{n}\sum_{\ell=1}^L\sum_{i\in\mathcal I_\ell}
\hat\psi_{i\ell}(\hat\theta)\hat\psi_{i\ell}(\hat\theta)'.
\]
The plug-in variance estimator is
\[
\hat V=(\hat G'\hat\Upsilon\hat G)^{-1}\hat G'\hat\Upsilon\hat\Psi\hat\Upsilon\hat G(\hat G'\hat\Upsilon\hat G)^{-1}.\]
\begin{remark}[Clustered data]\label{rem:cluster}
The results extend to samples of independent clusters of bounded size, such as households, by taking the cluster as the unit of observation and the sum of its members' scores as its moment. Theorem~\ref{thm:normality} then holds with the number of clusters in place of $n$ when the folds are formed by cluster, and $\Psi$ is estimated from within-cluster sums of $\hat\psi_{i\ell}$, as in the cluster-robust debiased machine learning framework of \citet{chiang2022multiway}.
\end{remark}
\begin{remark}[Testing the first-step correction]\label{rem:test}
Let $\hat\theta_2$ be the estimator that omits $\phi_1$ from $\psi$. Both estimators target $\theta_0$, so their difference tests whether the first-step estimation is negligible, not a hypothesis about $\theta_0$. By the first-step normal equations, the moment of $\hat\theta_2$ carries the drift $b_n=\sum_\ell(n_\ell/n)\int\alpha_{01}(\hat g_\ell-g_0)\,dF_0$, and the argument of Theorem~\ref{thm:normality} gives $\sqrt n(\hat\theta-\hat\theta_2)=-(G'\Upsilon G)^{-1}G'\Upsilon\{n^{-1/2}\sum_i\phi_1(W_i)-\sqrt n\,b_n\}+o_p(1)$. Under the null hypothesis $\sqrt n\,b_n=o_p(1)$, the difference is therefore asymptotically normal with a variance estimated from the cross-fitted first-step corrections, and the test rejects with probability approaching one when $\sqrt n\,b_n$ diverges.
\end{remark}

We now verify that Assumptions~\ref{ass:consistency} and \ref{ass:normality} are satisfied for the PLM estimator in equation~\eqref{eq:plm_estimator} under the following primitive conditions, and establish its asymptotic normality as a corollary of Theorem~\ref{thm:normality}.\par 

\begin{assumption}
\label{ass:plm}
The following conditions hold for the partially linear model:
\begin{enumerate}[(i)]
\item $B = \mathbb{E}[(X_3 - h_{0X}(X_2))(X_3 - h_{0X}(X_2))']$ exists and is nonsingular.
\item
$\mathbb E[|Y|^4]<\infty$,
$\mathbb E[\|X_3\|^4]<\infty$,
$\mathbb E[|g_0(X_1)|^4]<\infty$, and
$\mathbb E[\|\alpha_{01}(X_1)\|^4]<\infty$.\item For each fold $\ell$,
$\hat g_\ell\in\Delta_1$ and
$\hat\alpha_{01\ell}\in\Delta_1^p$
with probability approaching one. For each fold $\ell$,
\[
\|\hat{g}_\ell-g_0\|_4
+\|\hat{h}_\ell-h_0\|_4
+\|\hat{\alpha}_{01\ell}-\alpha_{01}\|_4
\overset{p}{\to}0,
\]
where $\|\cdot\|_4$ denotes the $L_4(F_0)$ norm,
with fitted functions held fixed when integrating.
\item For each fold $\ell$,
\begin{align*}
\sqrt{n}\left(\|\hat{g}_\ell-g_0\|_2^2
+\|\hat{h}_\ell-h_0\|_2^2\right)&=o_p(1),\\
\sqrt{n}\,
\|\hat{\alpha}_{01\ell}-\alpha_{01}\|_2
\|\hat{g}_\ell-g_0\|_2&=o_p(1).
\end{align*}
\end{enumerate}
\end{assumption}
\begin{corollary}[Asymptotic Normality for the PLM]\label{cor:plm_normality}
Under Assumption~\ref{ass:plm}, the estimator $\hat{\theta}$ in equation~\eqref{eq:plm_estimator} satisfies
\[\sqrt{n}(\hat{\theta} - \theta_0) \xrightarrow{d} \mathcal{N}(0, V),\]
where $V = B^{-1}\Psi B^{-1}$ and $\Psi = \mathbb{E}[\psi(W, g_0, h_0, \alpha_0, \theta_0)\psi(W, g_0, h_0, \alpha_0, \theta_0)']$. Moreover, the plug-in estimator $\hat{V} \overset{p}{\to} V$.
\end{corollary}
\subsection{Monte Carlo Simulations}
This section studies the finite-sample behavior of the proposed estimator
through Monte Carlo simulations for the partially linear model with an
ML-generated dependent variable. We begin with a first design in which
the covariates used to construct the generated outcome contain both the
controls and the regressor of interest. This design illustrates the
cancellation result in equation~\eqref{eq:lr_plm_main}: although the
parameter of interest is defined through an ML-generated outcome, the
locally robust moment eliminates the first-step regularization bias and
can be implemented using the observed outcome directly.

\paragraph{Estimators.} In both of the considered designs, the parameter of interest is the partially linear projection coefficient of a generated outcome $\tilde Y(g_0)$ on $X_3$ controlling for $X_2$, with $\nu\equiv X_3-h_{0X}(X_2)$. We compare three estimators. The infeasible oracle estimator uses the population nuisance functions and solves
\[
\hat\theta_{\mathrm{oracle}}
=
\Big(\frac1n\sum_{i=1}^n\nu_i^2\Big)^{-1}
\frac1n\sum_{i=1}^n\Big[\big\{\tilde Y_i(g_0)-h_{0Y}(X_{2i})\big\}\nu_i+\alpha_{01}(X_{1i})\big\{Y_i-g_0(X_{1i})\big\}\Big].
\]
The feasible locally robust estimator replaces each nuisance function by its cross-fitted estimate, evaluated only on held-out observations, and solves
\[
\hat\theta_{\mathrm{LR}}=\hat B^{-1}\hat C,
\qquad
\hat B=\frac1n\sum_{\ell=1}^L\sum_{i\in\mathcal I_\ell}\hat\nu_{i\ell}^2,
\qquad
\hat\nu_{i\ell}\equiv X_{3i}-\hat h_{0X,\ell}(X_{2i}),
\]
where
\[
\hat C=\frac1n\sum_{\ell=1}^L\sum_{i\in\mathcal I_\ell}\Big[\big\{\tilde Y_i(\hat g_\ell)-\hat h_{0Y,\ell}(X_{2i})\big\}\hat\nu_{i\ell}+\hat\alpha_{01,\ell}(X_{1i})\big\{Y_i-\hat g_\ell(X_{1i})\big\}\Big].
\]
The naive plug-in estimator $\hat\theta_{\mathrm{plugin}}=\hat B^{-1}\hat C_{\mathrm{plugin}}$ applies Robinson's partialling-out score directly to the cross-fitted generated outcome, so that $\hat C_{\mathrm{plugin}}$ retains only the first term of $\hat C$. This estimator ignores the regularization bias introduced by the first-step ML estimation of $g_0$. The two designs differ in the generated outcome $\tilde Y(g)$ and the first-step representer $\alpha_{01}$.\par

\paragraph{Design 1: generated outcome using all partially linear covariates.}
This simulation design mimics the following empirical scenario. A researcher seeks to estimate the effect of a binary variable of interest $X_3$ on an economic outcome, controlling for confounders $X_2$. The observed outcome $Y$ is noisy, and the researcher exploits additional high-dimensional features $Z$, such as unstructured text, images, or satellite-derived measurements, to construct a denoised prediction of the outcome using machine learning. The parameter of interest is the effect of $X_3$ on this ML-predicted outcome, controlling for $X_2$.

Specifically, we consider the data-generating process
\[Y = \theta_0 X_3+r_0(X_2)+s_0(Z)+\varepsilon, \qquad \varepsilon\sim\mathcal N(0,1),\]
where
\[r_0(X_2)=X_{21}^2+0.5X_{22}, \qquad s_0(Z)=0.1\sum_{j=1}^5 Z_j.\]
The confounders satisfy
\[X_{2j}\overset{\mathrm{i.i.d.}}{\sim}\mathcal N(0,1), \qquad j=1,\ldots,5,\]
the additional features satisfy
\[Z_j\overset{\mathrm{i.i.d.}}{\sim}\mathcal N(0,1), \qquad j=1,\ldots,10,\]
and the binary variable of interest is generated according to
\[X_3 = \mathbf 1\left(0.25X_{21}+0.25X_{22}+\eta>0\right), \qquad \eta\sim\mathcal N(0,1).\]
Thus $X_3$ is correlated with $X_2$, making the partialling-out step nontrivial. The true parameter is $\theta_0=1$.

In Design 1, the generated outcome is constructed using
\[X_1=(X_2,X_3,Z)\in\mathbb R^{16}.\]
Hence
\[\tilde Y \equiv g_0(X_1) \equiv \mathbb E[Y\mid X_1] = \theta_0X_3+r_0(X_2)+s_0(Z).\]
Since $X_2\subseteq X_1$ and $X_3\subseteq X_1$, the first-step Riesz representer simplifies to
\[\alpha_{01}(X_1) = X_3-h_{0X}(X_2), \qquad \alpha_{02}=0.\]
Therefore, the locally robust score reduces to Robinson's partialling-out moment evaluated at the observed outcome $Y$ rather than at the generated outcome $g(X_1)$, as established in equation \eqref{eq:lr_plm_main}. Thus, although the parameter of interest is defined through the ML-generated outcome $\tilde Y$, the locally robust estimator can be implemented without directly using $g$ in the final score.

The nuisance functions are estimated using $L=5$-fold cross-fitting. Let $\mathcal I_\ell$, $\ell=1,\ldots,L$, denote the folds. For each fold $\ell$, the nuisance estimators $\hat g_\ell$, $\hat h_{0Y,\ell}$, and $\hat h_{0X,\ell}$ are trained using only observations outside $\mathcal I_\ell$. Specifically, $\hat g_\ell$ is obtained by regressing $Y$ on $X_1=(X_2,X_3,Z)$ using LightGBM, while
\[\hat h_{0Y,\ell}(X_2) = \hat{\mathbb E}_\ell[Y\mid X_2], \qquad \hat h_{0X,\ell}(X_2) = \hat{\mathbb E}_\ell[X_3\mid X_2],\]
are obtained by regressing $Y$ and $X_3$ on $X_2$ using LightGBM. Estimating $\hat h_{0Y,\ell}$ directly from $Y$ rather than from $\hat g_\ell$ is valid since
\[\mathbb E[Y\mid X_2] = \mathbb E[g_0(X_1)\mid X_2]\]
by the law of iterated expectations.

In this design, $\tilde Y(g)=g(X_1)$ and $\alpha_{01}=\nu$, so the oracle and locally robust estimators use the observed outcome $Y_i$ in place of the generated outcome, with $h_{0Y}(X_2)=\mathbb E[Y\mid X_2]$, whereas the naive estimator uses $\hat g_\ell(X_{1i})$.\par

Table~\ref{tab:sims} reports bias, root mean square error (RMSE), and empirical coverage of nominal 95\% confidence intervals across $R=1,000$ Monte Carlo replications for sample sizes $n\in\{500,1000,2000,5000\}$. The locally robust estimator exhibits sound finite-sample performance. Its bias is negligible across all sample sizes, fluctuating around zero, and its RMSE is very similar to the oracle's. Coverage ranges from $0.934$ to $0.952$ and tracks that of the infeasible oracle. In contrast, the naive plug-in estimator exhibits a sizable bias that declines more slowly than $n^{-1/2}$, with coverage collapsing from $0.078$ to $0.003$.
\begin{table}[H]
\centering \resizebox{\textwidth}{!}{%
\begin{threeparttable} \begin{tabular}{l ccc ccc ccc} \hline & \multicolumn{3}{c}{Oracle} & \multicolumn{3}{c}{Locally Robust} & \multicolumn{3}{c}{Naive Plug-in} \\ \cmidrule(lr){2-4} \cmidrule(lr){5-7} \cmidrule(lr){8-10}
$n$ & Bias & RMSE & Coverage & Bias & RMSE & Coverage & Bias & RMSE & Coverage \\
\hline\hline
500 & 0.003 & 0.095 & 0.936 & 0.010 & 0.111 & 0.951 & -0.277 & 0.304 & 0.078 \\
1,000 & 0.000 & 0.069 & 0.947 & -0.004 & 0.077 & 0.943 & -0.209 & 0.224 & 0.035 \\
2,000 & 0.001 & 0.048 & 0.939 & -0.002 & 0.053 & 0.934 & -0.147 & 0.157 & 0.028 \\
5,000 & -0.002 & 0.029 & 0.956 & -0.006 & 0.031 & 0.952 & -0.100 & 0.105 & 0.003 \\
\hline
\end{tabular}
\begin{tablenotes}
\item Results based on $R=1,000$ Monte Carlo replications. The true parameter is $\theta_0=1$. Nominal coverage is $0.95$. All nuisance functions are estimated via LightGBM with $L=5$-fold cross-fitting.
\end{tablenotes}
\end{threeparttable}}
\caption{Finite-sample performance of the oracle, locally robust, and naive plug-in estimators in Design 1.}
\label{tab:sims}
\end{table}

\paragraph{Design 2: residualized generated outcome with a nested second step.}
This design mimics a second empirical scenario. A researcher is interested in the association of a binary variable of interest $X_3$, say sex, with lifetime income, but the outcome is first purged of confounding variation, such as year and cohort dummies, before projecting the residualized outcome on $X_3$ controlling for $X_2$. The parameter of interest is therefore the partially linear projection coefficient of the residualized outcome, as in Remark~\ref{rem:plm_projection}. This is also the general case of our framework, in which the first-step correction does not collapse and the second-step nuisance depends on the first step.\par 
We maintain the outcome equation of Design 1 and the distributions of the confounders $X_2$, the shock $\eta$ and the treatment $X_3$, with $\beta_0=1$, and modify the following features:
\[Z_1 = \rho\,\eta+\sqrt{1-\rho^2}\,\omega, \quad \rho=0.6, \qquad s_0(Z)=0.1\sum_{j=2}^{6}Z_j, \qquad \dim(Z)=50,\]
\[X_1=(X_{21},X_{22},Z)\in\mathbb R^{52}, \qquad r_0(X_2)=X_{21}^2+0.5X_{22}+X_{23},\]
with $\omega\sim\mathcal N(0,1)$ and $Z_j\overset{\mathrm{i.i.d.}}{\sim}\mathcal N(0,1)$ for $j=2,\ldots,50$. The proxy $Z_1$ makes $X_1$ partially informative about the treatment's identifying variation $\nu\equiv X_3-h_{0X}(X_2)$, so that $\alpha_{01}$ neither vanishes nor equals $\nu$; excluding $Z_1$ from $s_0$ keeps it free of any direct effect on the outcome; the higher dimension prevents the regularization bias in $\hat g$ from vanishing too quickly; and excluding $X_{23}$ from $X_1$ makes the second step nontrivial.

The generated outcome is the observed outcome residualized of the variation explained by $X_1$,
\[\tilde Y \equiv Y-g_0(X_1), \qquad g_0(X_1)\equiv\mathbb E[Y\mid X_1],\]
as in Remark \ref{rem:resid}, so that $\alpha_{02}=0$ and $\alpha_{01}(X_1)=-\mathbb E[\nu\mid X_1]$, and
\[\psi(W,g_0,h_0,\alpha_0,\theta) = \left(\tilde Y-h_{0Y}(X_2)-\theta\nu\right)\nu + \alpha_{01}(X_1)\left(Y-g_0(X_1)\right).\]
Two features distinguish this design from Design 1. First, $\alpha_{01}$ must be estimated, since $X_3\not\subseteq X_1$. Second, $h_{0Y}(X_2)=X_{23}\neq0$ cannot be recovered from $Y$ alone, so the second step must be estimated from the first. This makes Design~2 a nested case of the framework in Section~\ref{sec:set}, in which estimation error from the first step propagates into the second-step nuisance and is handled through the joint remainder conditions of our asymptotic theory. The estimand is
\[\theta_0 = \beta_0\left(1-R^2_{\nu\mid X_1}\right), \qquad R^2_{\nu\mid X_1} \equiv \frac{\mathbb E[\alpha_{01}^2]}{\mathbb E[\nu^2]},\]
which admits a closed form in this design and equals $0.23$ for $\rho=0.6$, so that $\theta_0=0.77$. Unlike Design 1, the partially linear projection coefficient differs from $\beta_0$: residualizing removes the component of the identifying variation $\nu$ that is explained by $X_1$.

Nuisances are estimated as in Design 1, with $L=5$-fold cross-fitting and two differences. The Riesz representer is estimated by $\hat\alpha_{01,\ell}=-\big(\hat{\mathbb E}_\ell[X_3\mid X_1]-\hat{\mathbb E}_\ell[\hat h_{0X,\ell}(X_2)\mid X_1]\big)$, with both conditional expectations of the binary $X_3$ obtained by a logistic objective and an inner cross-fit for the values of $\hat h_{0X}$ entering the projection; projecting onto $X_1$ makes the representer $X_1$-measurable by construction, without imposing any condition on the data-generating process. The second-step nuisance $\hat h_{0Y,\ell}$ is obtained by regressing $Y-\hat g$ on $X_2$, again using an inner cross-fit so that the first-step fits are out of sample. With $\tilde Y(g)=Y-g(X_1)$ and these estimates, the three estimators follow directly from the general definitions above.\par 
Table~\ref{tab:sims2} reports the same statistics for Design 2. The locally robust estimator again exhibits sound finite-sample performance, with negligible bias across all sample sizes and RMSE approaching that of the oracle. However, the coverage ranges from $0.918$ to $0.934$, some two percentage points below the oracle's. Two sources could account for this shortfall: the interval may be off-center, or it may be the wrong width. To tell them apart, note that coverage depends on both through a single expression. Let $sd$ denote the Monte Carlo standard deviation of the estimator and write $\lambda \equiv \text{mean}(\hat{se})/sd$ and $b\equiv\text{bias}/sd$. Standardizing $|\hat\theta-\theta_0|\le 1.96\,\hat{se}$ by $sd$ and using the asymptotic normality established in Theorem~\ref{thm:normality}, so that $(\hat\theta-\theta_0)/sd$ is approximately $\mathcal N(b,1)$, gives
\[\text{Coverage} \approx \Phi(1.96\lambda-b)-\Phi(-1.96\lambda-b),\]
which equals the nominal $0.95$ when $\lambda=1$ and $b=0$. The two departures can therefore be switched off separately.\par 
\begin{table}[H]
\centering \resizebox{\textwidth}{!}{%
\begin{threeparttable} \begin{tabular}{l ccc ccc ccc} \hline & \multicolumn{3}{c}{Oracle} & \multicolumn{3}{c}{Locally Robust} & \multicolumn{3}{c}{Naive Plug-in} \\ \cmidrule(lr){2-4} \cmidrule(lr){5-7} \cmidrule(lr){8-10}
$n$ & Bias & RMSE & Coverage & Bias & RMSE & Coverage & Bias & RMSE & Coverage \\
\hline\hline
500 & 0.002 & 0.099 & 0.946 & 0.023 & 0.138 & 0.924 & 0.118 & 0.171 & 0.829 \\
1,000 & 0.001 & 0.071 & 0.947 & 0.000 & 0.089 & 0.924 & 0.094 & 0.123 & 0.792 \\
2,000 & -0.001 & 0.050 & 0.950 & -0.006 & 0.061 & 0.918 & 0.073 & 0.093 & 0.695 \\
5,000 & -0.002 & 0.031 & 0.952 & -0.003 & 0.035 & 0.934 & 0.055 & 0.064 & 0.582 \\
\hline
\end{tabular}
\begin{tablenotes}
\item Results based on $R=1,000$ Monte Carlo replications. The true parameter is $\theta_0=\beta_0(1-R^2_{\nu\mid X_1})=0.77$. Nominal coverage is $0.95$. All nuisance functions are estimated via LightGBM with $L=5$-fold cross-fitting.
\end{tablenotes}
\end{threeparttable}}
\caption{Finite-sample performance of the oracle, locally robust, and naive plug-in estimators in Design 2.}
\label{tab:sims2}
\end{table}
Table~\ref{tab:coverage_decomp} reports this decomposition for both designs. The standardized bias is negligible at every sample size, so that setting $b=0$ leaves coverage essentially unchanged; setting $\lambda=1$ instead, and leaving the bias as it is, restores coverage of $0.946$--$0.950$ in every cell. Predicted coverage, which treats the standard error as fixed and the estimator as normal, lies within 0.004 of observed coverage in every cell. The shortfall in Design 2 is therefore consistent with an understatement of the standard error on average, rather than with bias in the estimator. The pattern is consistent with higher-order variability from nuisance estimation that is not captured by the first-order asymptotic variance approximation. Design 1 estimates three nuisance functions, and $\lambda$ is close to unity throughout, so that the cost of width is at most 1.2 percentage points and coverage is essentially nominal. Design 2 estimates five, including a Riesz representer and a second-step nuisance that is itself constructed from the first-step fit, and, up to $n=2{,}000$, $\lambda$ falls to roughly $0.90$, with a corresponding cost of 2.3 to 3.1 percentage points. The gap closes as the sample grows ($\lambda$ rises to $0.947$ at $n=5,000$), consistent with the omitted contribution being of smaller order.
\begin{table}[H]
\centering 
\begin{threeparttable} \begin{tabular}{l cc ccc c} \hline
$n$ & $\lambda$ & $b$ & Observed & Predicted & At $\lambda=1$ & Cost of width \\
\hline\hline
\multicolumn{7}{l}{\textit{Design 1}} \\\hline
500 & 0.994 & 0.094 & 0.951 & 0.947 & 0.949 & 0.002 \\
1,000 & 0.956 & -0.054 & 0.943 & 0.939 & 0.950 & 0.011 \\
2,000 & 0.953 & -0.032 & 0.934 & 0.938 & 0.950 & 0.012 \\
5,000 & 1.028 & -0.197 & 0.952 & 0.952 & 0.946 & -0.006 \\
\hline
\multicolumn{7}{l}{\textit{Design 2}} \\\hline
500 & 0.906 & 0.171 & 0.924 & 0.920 & 0.947 & 0.027 \\
1,000 & 0.915 & 0.002 & 0.924 & 0.927 & 0.950 & 0.023 \\
2,000 & 0.893 & -0.092 & 0.918 & 0.919 & 0.949 & 0.031 \\
5,000 & 0.947 & -0.091 & 0.934 & 0.936 & 0.949 & 0.014 \\
\hline
\end{tabular}
\begin{tablenotes}
\item Decomposition of the coverage attained by the locally robust estimator, based on $R=1,000$ Monte Carlo replications. Predicted coverage evaluates $\Phi(1.96\lambda-b)-\Phi(-1.96\lambda-b)$ at the measured $\lambda$ and $b$. The column at $\lambda=1$ sets the interval to the estimator's sampling standard deviation, leaving the bias unchanged; it is a counterfactual, not a feasible procedure. The final column is the difference between the two.
\end{tablenotes}
\end{threeparttable}
\caption{Decomposition of coverage into interval width and bias.}
\label{tab:coverage_decomp}
\end{table}
Monte Carlo simulations illustrate that our framework delivers valid inference with a machine-learning generated dependent variable.  The locally robust estimator is essentially unbiased in both designs, its root mean square error converges on that of the infeasible oracle, and its confidence intervals attain close to nominal coverage, with the residual shortfall consistent with an understatement of the standard error rather than with bias. The naive plug-in estimator, by contrast, carries a bias that vanishes too slowly for valid inference, and its coverage deteriorates as the sample grows. The framework thus offers a practical toolkit for valid inference in the presence of outcomes predicted by ML.
\section{Empirical Application}
\label{sec:empirical}
We apply the life-cycle correction of Section~\ref{sec:iop} to measure inequality of opportunity in Spain, using the 2023 Encuesta de Condiciones de Vida (ECV), the Spanish component of the European Union Statistics on Income and Living Conditions (EU-SILC). The survey records a single year of income for individuals observed at different ages, so measured IOp is even more exposed to life-cycle bias than with longitudinal data, where averaging income over several years mitigates it. To correct it, we shift each individual from the expected income of their type at their observed age to the expected income of the same type over the working life, as in \eqref{eq:iop_purge}. Because these type-specific age--income profiles are estimated by machine learning, corrected income is a generated outcome, and relative IOp falls within our framework. We measure relative IOp with the variance of logs and the mean log deviation (MLD), two indices widely used in the IOp literature \citep{ferreira2011measurement}, whose locally robust moments are given in Corollary~\ref{cor:iop}. We compare these estimates with two estimators that treat corrected income as observed, a plug-in estimator and one that corrects only for the estimation of the type averages, and with the quadratic adjustment and no adjustment used in the literature.\par
The 2023 wave of the ECV includes a module on the intergenerational transmission of disadvantage, which asks adults about their family circumstances when they were around 14 years old. Our outcome, $Y$, is log household disposable income divided by the modified OECD equivalence scale. We apply the life-cycle correction to this equivalized measure, so that it removes the life-cycle variation in household needs as well as in income. Age, $A$, is measured at the end of the income reference year, and the module covers adults aged 25 to 59. We use eight circumstances, $C$: sex, own and parents' countries of birth, parental education, the mother's activity status, and the household's financial situation and ability to afford holidays at around age 14. Excluding the 0.3\% of respondents with nonpositive income and those with missing circumstances leaves 27,301 individuals. Following \citet{brunori2025inequality}, our main sample keeps household heads and their spouses, whose equivalized income reflects their own resources rather than those of their parents. The final sample contains 22,491 individuals in 15,093 households.\par
Because machine learning estimates the age--income profiles of many types flexibly, our correction can accommodate a wide set of circumstances. The choice of circumstances nonetheless involves a trade-off. Ex-ante estimates are lower bounds that rise with the set of circumstances \citep{ferreira2011measurement}, so a richer set tightens the bound. At the same time, it partitions the sample into more types, so that the age distribution within each type becomes harder to estimate and the weights $w_{\mathcal{R}}$ entering the first-step correction may become extreme. When the set of circumstances is large, estimating the Riesz representer automatically \citep{chernozhukov2022locally}, without inverting estimated age probabilities, is a natural alternative, although its performance depends on the richness of the dictionary and it may be less stable for indices on income levels, such as the MLD, whose correction involves exponentiated income. The eight circumstances listed above are the six whose exclusion most reduces the out-of-sample fit of income, together with own and mother's country of birth, which complete the information on countries of birth. With all 15 circumstances in the module, the locally robust estimates are essentially unchanged.\par
The correction requires choosing the reference ages $\mathcal{R}$ in \eqref{eq:iop_ref}. It makes individuals observed at different ages comparable, but the reference ages still determine the stage of the life cycle at which IOp is measured. We use the whole working life, ages 25 to 59, as reference ages, with equal weights, $p_{\mathcal{R}}(b)=1/35$, so that the reference income \eqref{eq:iop_ref} is the average annual income of the type over the working life, regardless of the age structure of the population. Assuming that age profiles are stable across cohorts within types, the type averages then measure the opportunities of each type over the working life. Many types are rarely or never observed at some ages, especially the youngest, so we estimate the age distribution within types with a multinomial logit over ages that leaves the common age distribution unrestricted and lets each type tilt it smoothly, linearly and quadratically in age, through the main effects of its circumstances, with a penalty on the tilts.\par
We estimate the age--income profiles of the types, $g_0(a,c)=\mathbb{E}[Y\mid A=a,C=c]$ in \eqref{eq:iop_lc}, with three learners tuned by cross-validation: LightGBM, lasso on circumstance indicators, age splines and their interactions, and random forest. The remaining nuisances are the type averages $h_0$; the conditional means of corrected income \eqref{eq:iop_purge} given age and type, and given type alone, which enter $\alpha_{01}$ in Proposition~\ref{prop:iop}; and the age distribution within types $p_0(b\mid c)$, which enters the weights $w_{\mathcal{R}}$. We estimate the type averages by LightGBM tuned by cross-validation, which also selects between LightGBM, lasso and their average, and the conditional means entering $\alpha_{01}$ by LightGBM with a fixed, strongly regularized structure and the number of trees chosen by early stopping. We tune the penalty of the age-distribution model described above by cross-validation and normalize the weights in each fold so that each reference age receives exactly its weight $p_{\mathcal{R}}(b)$. All nuisances are cross-fitted with five folds formed by household, and the generated outcomes used to train the second steps are themselves cross-fitted within each training sample. We estimate $\theta_0$ by the sample analogue of the moments in Proposition~\ref{prop:iop} (LR). We compare it with two estimators that use the same nuisances but treat corrected income as observed: the plug-in estimator (naive), and the estimator that is locally robust to the type averages alone (LR$_2$) \citep{chernozhukov2022locally}. We report medians over five cross-fitting splits. Standard errors are clustered by household, as in Remark~\ref{rem:cluster}.\par 
According to Table~\ref{tab:iop_main}, treating corrected income as observed overstates inequality of opportunity. The naive and LR$_2$ estimates, which do so, attribute between 9.2\% and 9.9\% of the variance of log income and between 10.0\% and 11.0\% of its MLD to circumstances, depending on the learner. Once corrected income is treated as a generated outcome, the LR estimates fall to between 8.4\% and 8.6\% and between 9.3\% and 9.6\%. With the heterogeneous correction, correcting only for the estimation of the type averages, as LR$_2$ does, changes relative IOp by at most 0.15 percentage points. Correcting also for the estimation of the age--income profiles that generate corrected income lowers it by 1.3 to 1.5 points with LightGBM, or 13\% of the LR$_2$ estimate, and by 0.8 to 1.2 points with the lasso and forests. The test of Remark~\ref{rem:test} rejects that this correction is negligible at the 5\% level with LightGBM for both indices and at the 10\% level with forests for the variance of logs; with the lasso, the correction has the same sign but is smaller, and the test does not reject. It also brings the learners into agreement: the LR estimates differ across learners by about 0.2 points, against 0.7 and 0.8 points for LR$_2$.\par
\begin{table}[t]
\centering
\small
\begin{threeparttable}
\caption{Relative Inequality of Opportunity in Spain}
\label{tab:iop_main}
\begin{tabular*}{\textwidth}{@{\extracolsep{\fill}}l*{5}{c}}
\toprule
 & \multicolumn{2}{c}{Benchmark} & \multicolumn{3}{c}{Heterogeneous correction} \\
\cmidrule(lr){2-3}\cmidrule(lr){4-6}
 & None & Quadratic & LightGBM & Lasso & Forest \\
\midrule
\multicolumn{6}{l}{\textit{Variance of logs}} \\ \hline
Naive              & 9.57   & 9.75   & 9.85   & 9.20   & 9.64   \\
LR$_2$             & 9.79   & 9.89   & 9.89   & 9.20   & 9.69   \\
                   & (0.56) & (0.56) & (0.57) & (0.54) & (0.56) \\
LR                 &        & 9.89   & 8.60   & 8.44   & 8.64   \\
                   &        & (0.56) & (0.85) & (0.79) & (0.87) \\
LR $-$ LR$_2$      &        & 0.00   & $-$1.29$^{**}$ & $-$0.79 & $-$1.06$^{*}$ \\
                   &        & (0.01) & (0.60) & (0.57) & (0.62) \\
\addlinespace \midrule
\multicolumn{6}{l}{\textit{Mean log deviation}} \\ \hline
Naive              & 10.39  & 10.77  & 10.84  & 9.98   & 10.48  \\
LR$_2$             & 10.57  & 10.82  & 10.96  & 10.12  & 10.61  \\
                   & (0.45) & (0.46) & (0.46) & (0.45) & (0.45) \\
LR                 &        & 10.82  & 9.56   & 9.33   & 9.49   \\
                   &        & (0.46) & (0.86) & (0.78) & (0.86) \\
LR $-$ LR$_2$      &        & 0.00   & $-$1.45$^{**}$ & $-$0.88 & $-$1.16 \\
                   &        & (0.03) & (0.70) & (0.63) & (0.71) \\
\bottomrule
\end{tabular*}
\begin{tablenotes}
\scriptsize
\item Notes: Relative IOp (\%), ECV 2023, household heads and spouses aged 25--59, $n=22,491$. Eight circumstances. \textit{None}: observed income. \textit{Quadratic}: quadratic age profile, estimated by regressing log income on age and age squared, removed. Last three columns: correction \eqref{eq:iop_purge} with reference ages 25--59 and the indicated first-step learner; medians over five cross-fitting splits, with split-adjusted standard errors \citep{chernozhukov2018double}, taken separately for each row. \textit{Naive}: plug-in estimator, without correction for regularization bias; standard errors not reported. LR$_2$: locally robust to the estimation of the type averages \citep{chernozhukov2022locally}, treating corrected income as observed. LR: locally robust to both steps (Proposition~\ref{prop:iop}). Without a first step, LR and LR$_2$ coincide. All estimators use the same cross-fitted nuisances. Unweighted; household-clustered standard errors in parentheses. $^{*}$, $^{**}$, $^{***}$: the test of Remark~\ref{rem:test} rejects at 10\%, 5\%, 1\%.
\end{tablenotes}
\end{threeparttable}
\end{table}
The life-cycle correction is also sizable relative to the adjustments used in the literature. Without adjustment, circumstances account for 9.8\% of the variance of log income and 10.6\% of its MLD, and the quadratic adjustment raises these figures only to 9.9\% and 10.8\%. The LR estimates lie 1.2 to 1.5 percentage points below the quadratic adjustment and 1.0 to 1.4 points below the unadjusted estimates, that is, 10\% to 15\% lower. This gap appears only once the generated outcome is accounted for. With LightGBM and forests, the LR$_2$ estimates are within 0.2 points of the quadratic adjustment, so a researcher who replaced the quadratic adjustment with type-specific profiles but treated corrected income as observed would conclude that the choice of adjustment barely matters..\par
Our estimates are consistent with existing evidence for Spain, once differences in data and methods are taken into account. The unadjusted estimates are of the same order as those of \citet{palomino2019channels}. Using EU-SILC with incomes of 2004 and 2010, the authors find that circumstances account for 12.1\% and 12.5\% of the MLD of equivalized income, and that the $R^2$ of log income on circumstances, the parametric counterpart of our variance-of-logs share, is 0.098 and 0.096. The locally robust estimates, 8.4--8.6\% and 9.3--9.6\%, lie below these figures and well below those of studies with richer circumstance sets or broader samples. \citet{cabrera2021inequality} find 17.7\% of the MLD with a 2017 survey that adds school type, siblings and parental cultural capital. The forest estimates of the GEOM database \citep{ferreira2026geom} give 16.8\% for EU-SILC 2019, in a sample that includes all adults aged 18 and over. These studies address the life cycle in different ways from ours: \citet{cabrera2021inequality} restrict ages to 25--60, and GEOM removes a common quadratic profile. \citet{filauro2023evolution}, who instead treat age as a circumstance, find that it is the largest single contributor to IOp in Spanish earnings, at about 40\% of its MLD, which underscores the importance of accounting for the life cycle. These comparisons are scale checks rather than replications, since years, surveys, sample definitions, circumstances, income definitions and estimators differ, and ex-ante estimates are lower bounds that rise with the set of circumstances.\par 
\section{Conclusions}\label{sec:conc}
In this paper, we provide a framework for locally robust inference with outcomes predicted by machine learning. We consider a three-step setting in which, in the first step, an outcome is generated by ML; in the second step, a nuisance function is constructed using nonparametric methods that involve the generated outcome; and in the third step, the parameter of interest is identified through a GMM criterion. To attain valid inference, we construct a moment that is Neyman-orthogonal to the first- and second-step nuisances, exploiting the idea of downstream local robustness: after correcting for the second step, the first-step correction is obtained by differentiating the augmented moment with respect to the first step, holding the second-step nuisance fixed. By replacing restrictive parametric first steps with flexible machine learning methods, researchers can reduce sensitivity to functional-form misspecification, while our debiased moment condition corrects for the first-order regularization bias that ML estimation introduces, enabling valid inference under the stated conditions.\par
We establish the consistency and asymptotic normality of the resulting estimator and illustrate its finite-sample behavior through Monte Carlo simulations. The locally robust estimator exhibits small bias and coverage close to nominal levels, whereas the naive estimator that treats the ML-generated outcome as observed can display substantial bias and severe undercoverage. We apply the method to measure inequality of opportunity in Spain, where incomes observed at different ages are subject to life-cycle bias. Removing this bias lowers inequality of opportunity by 10\% to 14\% relative to raw income, We establish the consistency and asymptotic normality of the resulting estimator and illustrate its finite-sample behavior through Monte Carlo simulations. The locally robust estimator exhibits small bias and coverage close to nominal levels, whereas the naive estimator that treats the ML-generated outcome as observed can display substantial bias and severe undercoverage. We apply the method to measure inequality of opportunity in Spain, where incomes observed at different ages are subject to life-cycle bias. Removing this bias lowers inequality of opportunity by 10\% to 14\% relative to raw income, whereas treating the corrected incomes as observed overstates it by up to 15\%. \par  
\bibliography{cites}\clearpage
\begin{appendices}
\section{Proofs}\label{sec:proofs}
\textbf{Proof of Theorem \ref{thm:lr}:} Fix $\theta\in\Theta$. By Assumption \ref{ass:lr}, the partial
derivative functionals
\[\dot m_g(\cdot;\theta):\Delta_1\to\mathbb{R}^q,\qquad\dot m_h(\cdot;\theta):\Delta_2\to\mathbb{R}^q \]
are continuous and linear. Moreover, since $g\mapsto S(g)$ is Hadamard differentiable and $h_0(g)$ is the $L_2$ projection of $S(g)$ onto $\Delta_2$, the map $g\mapsto h_0(g)$ is Hadamard differentiable, with derivative
\[\dot h_g(\delta_1)\equiv D_gh_0(g_0)[\delta_1]\in\Delta_2\] 
characterized by
\[\mathbb{E}\left[\delta_2(X_2) \left\{ D_gS(g_0)[\delta_1] - \dot h_g(\delta_1)(X_2)\right\}\right]=0
\quad
\text{for all } \delta_2\in\Delta_2.
\]
Thus
\[
\dot m_{g,\mathrm{tot}}(\delta_1;\theta)
=
\dot m_g(\delta_1;\theta)
+
\dot m_h(\dot h_g(\delta_1);\theta)
\]
is also a continuous linear functional on $\Delta_1$. Since
$\Delta_1$ and $\Delta_2$ are closed linear subspaces of
$L_2(X_1)$ and $L_2(X_2)$, they are Hilbert spaces. Therefore, by the
Riesz representation theorem, there exist unique
$\alpha_{01}(\cdot;\theta)\in\Delta_1$ and
$\alpha_{02}(\cdot;\theta)\in\Delta_2$ satisfying
\eqref{eq:riesz1}--\eqref{eq:riesz2}.

We next show that the debiased moment has the same mean as the
original identifying moment at the truth. Since
$\alpha_{01}(\cdot;\theta_0)\in\Delta_1$ and $g_0$ satisfies the
first-step projection condition,
\[
\mathbb{E}\left[
\alpha_{01}(X_1;\theta_0)\{Y-g_0(X_1)\}
\right]=0.
\]
Similarly, since $\alpha_{02}(\cdot;\theta_0)\in\Delta_2$ and
$h_0=h_0(g_0)$ is the second-step projection of $S(g_0)$,
\[
\mathbb{E}\left[
\alpha_{02}(X_2;\theta_0)'
\{S(g_0)-h_0(X_2)\}
\right]=0.
\]
Consequently,
\[
\mathbb{E}[\psi(W,g_0,h_0,\alpha_0,\theta_0)]
=
\mathbb{E}[m(W,g_0,h_0,\theta_0)].
\]
Thus, if $\mathbb{E}[m(W,g_0,h_0,\theta_0)]=0$, then
\[
\mathbb{E}[\psi(W,g_0,h_0,\alpha_0,\theta_0)]=0.
\]

It remains to establish Neyman orthogonality. Let
\[
g_\tau=g_0+\tau\delta_1+o(\tau),
\qquad
h_\tau=h_0+\tau\delta_2+o(\tau),
\]
with $\delta_1\in\Delta_1$ and $\delta_2\in\Delta_2$. By Hadamard
differentiability of $\bar m$,
\[
\frac{d}{d\tau}
\mathbb{E}[m(W,g_\tau,h_\tau,\theta)]
\bigg|_{\tau=0}
=
\dot m_g(\delta_1;\theta)+\dot m_h(\delta_2;\theta).
\]
Using the definition of the total derivative representer,
\[
\dot m_g(\delta_1;\theta)
=
\mathbb{E}[\alpha_{01}(X_1;\theta)\delta_1(X_1)]
-
\dot m_h(\dot h_g(\delta_1);\theta).
\]
Using the definition of $\alpha_{02}$,
\[
\dot m_h(\delta_2;\theta)
=
\mathbb{E}[\alpha_{02}(X_2;\theta)'\delta_2(X_2)]
\]
and
\[
\dot m_h(\dot h_g(\delta_1);\theta)
=
\mathbb{E}\left[
\alpha_{02}(X_2;\theta)'
\dot h_g(\delta_1)(X_2)
\right].
\]
Therefore,
\begin{align*}
\frac{d}{d\tau}
\mathbb{E}[m(W,g_\tau,h_\tau,\theta)]
\bigg|_{\tau=0}
&=
\mathbb{E}[\alpha_{01}(X_1;\theta)\delta_1(X_1)]
\\
&\quad
-
\mathbb{E}\left[
\alpha_{02}(X_2;\theta)'
\dot h_g(\delta_1)(X_2)
\right]
+
\mathbb{E}[\alpha_{02}(X_2;\theta)'\delta_2(X_2)].
\end{align*}

The derivative of the first correction term is
\[
\frac{d}{d\tau}
\mathbb{E}\left[
\alpha_{01}(X_1;\theta)\{Y-g_\tau(X_1)\}
\right]\bigg|_{\tau=0}
=
-
\mathbb{E}[\alpha_{01}(X_1;\theta)\delta_1(X_1)].
\]
The derivative of the second correction term is
\begin{align*}
&\frac{d}{d\tau}
\mathbb{E}\left[
\alpha_{02}(X_2;\theta)'
\{S(g_\tau)-h_\tau(X_2)\}
\right]\bigg|_{\tau=0}
\\
&\quad =
\mathbb{E}\left[
\alpha_{02}(X_2;\theta)'D_gS(g_0)[\delta_1]
\right]
-
\mathbb{E}\left[
\alpha_{02}(X_2;\theta)'\delta_2(X_2)
\right].
\end{align*}
Because $\dot h_g(\delta_1)$ is the $L_2$ projection of
$D_gS(g_0)[\delta_1]$ onto $\Delta_2$ and
$\alpha_{02}(\cdot;\theta)\in\Delta_2$,
\[
\mathbb{E}\left[
\alpha_{02}(X_2;\theta)'D_gS(g_0)[\delta_1]
\right]
=
\mathbb{E}\left[
\alpha_{02}(X_2;\theta)'
\dot h_g(\delta_1)(X_2)
\right].
\]
Combining the derivatives of the original moment and the two correction
terms, all terms cancel:
\[
\frac{d}{d\tau}
\mathbb{E}\left[
\psi(W,g_\tau,h_\tau,\alpha_0,\theta)
\right]\bigg|_{\tau=0}
=0.
\]
This proves Neyman orthogonality and completes the proof.
$\square$.  

\bigskip \noindent
\textbf{Proof of Proposition \ref{prop:plm}:}
By the definitions of $h_{0Y}$ and $h_{0X}$,
\[
\mathbb E[g_0(X_1)-h_{0Y}(X_2)\mid X_2]=0,
\qquad
\mathbb E[\nu\mid X_2]=0.
\]
Hence the Robinson moment is orthogonal to both components of
$h_0$, so $\alpha_{02}=0$. For every $\delta_1\in\Delta_1$,
the total first-step derivative is
\[
\mathbb E[\delta_1(X_1)\nu]
=
\mathbb E[
\delta_1(X_1)\Pi_{\Delta_1}\nu
],
\]
which identifies $\alpha_{01}=\Pi_{\Delta_1}\nu$.
Since each component of $\alpha_{01}$ belongs to $\Delta_1$,
the first-step normal equations also give
\[
\mathbb E[
\alpha_{01}(X_1)\{Y-g_0(X_1)\}
]=0.
\]
Theorem~\ref{thm:lr} therefore yields
\eqref{eq:lr_plm_general_main}. When
$\Delta_1=L_2(X_1)$, the projection is conditional expectation.
If $X_2$ and $X_3$ are contained in $X_1$, then
$\alpha_{01}=\nu$ and iterated expectations give
$h_{0Y}(X_2)=\mathbb E[Y\mid X_2]$. Substitution yields
\eqref{eq:lr_plm_main}.
$\square$

\bigskip \noindent
\textbf{Proof of Proposition \ref{prop:app}:}
By \eqref{eq:prod_h0},
\[
\omega_1
=
h_0(X_{21})
+
\mathbb{E}[\omega_1\mid X_{11}].
\]
Using \eqref{eq:prod_struct} and \eqref{eq:prod_ar1_struct},
\begin{align*}
Y_2-\theta_0 h_0(X_{21})
&=
\tilde g_{02}(K_2,L_2)
+
\theta_0\omega_1
+
\xi_2
+
U_2
-
\theta_0 h_0(X_{21})
\\
&=
\tilde g_{02}(K_2,L_2)
+
\theta_0
\mathbb{E}[\omega_1\mid X_{11}]
+
\xi_2
+
U_2.
\end{align*}
Since
\[
\tilde g_{02}(K_2,L_2)\in L_2(X_{12})
\qquad\text{and}\qquad
\mathbb{E}[\omega_1\mid X_{11}]\in L_2(X_{11}),
\]
both functions belong to $\mathcal H$. By construction,
$Z_0=R-\Pi_{\mathcal H}R$ is orthogonal to $\mathcal H$, and therefore
\[
\mathbb{E}\left[
Z_0\tilde g_{02}(K_2,L_2)
\right]
=
0,
\qquad
\mathbb{E}\left[
Z_0\mathbb{E}[\omega_1\mid X_{11}]
\right]
=
0.
\]
Moreover, $Z_0$ is $\mathcal F_1$-measurable, so
\eqref{eq:prod_timing2} implies
\[
\mathbb E[Z_0(\xi_2+U_2)]
=
\mathbb E\!\left[
Z_0\mathbb E[\xi_2+U_2\mid\mathcal F_1]
\right]
=0.
\]
Hence
\[
\mathbb{E}\left[
Z_0
\left\{
Y_2-\theta_0 h_0(X_{21})
\right\}
\right]
=
0.
\]
If $\mathbb{E}[Z_0h_0(X_{21})]\neq0$, solving this moment condition
gives
\[
\theta_0
=
\frac{\mathbb{E}[Z_0Y_2]}
{\mathbb{E}[Z_0h_0(X_{21})]}.
\]

It remains to characterize existence of a relevant instrument in
$\mathcal Z$. Let
\[
M_{\mathcal H}
\equiv
I-\Pi_{\mathcal H}.
\]
For any $r\in L_2(X_{21})$, the associated instrument is
\[
Z_r=M_{\mathcal H}r(X_{21}).
\]
Since $Z_r\perp\mathcal H$,
\[
\mathbb{E}[Z_rh_0(X_{21})]
=
\mathbb{E}\left[
Z_r\,M_{\mathcal H}h_0(X_{21})
\right].
\]
If $h_0(X_{21})\in\mathcal H$, then
$M_{\mathcal H}h_0(X_{21})=0$, so every instrument in
$\mathcal Z$ is irrelevant. Conversely, if
$h_0(X_{21})\notin\mathcal H$, choose
$r(X_{21})=h_0(X_{21})$. Then
\[
Z_0^*
=
M_{\mathcal H}h_0(X_{21})
=
h_0(X_{21})-\Pi_{\mathcal H}h_0(X_{21}),
\]
and
\[
\mathbb{E}[Z_0^*h_0(X_{21})]
=
\mathbb{E}[(Z_0^*)^2]
>
0.
\]
Thus a relevant instrument in $\mathcal Z$ exists if and only if
$h_0(X_{21})\notin\mathcal H$.
$\square$

\bigskip \noindent
\textbf{Proof of Proposition \ref{prop:prod_lr}:}
The identifying moment is
\[
m(W,g,h,\theta)
=
\{R-q(X_{11},X_{12})\}
\{Y_2-\theta h(X_{21})\}.
\]

We first consider the second-step nuisance. For
$h_\tau=h_0+\tau\delta_2$,
\[
\frac{d}{d\tau}
\mathbb E[
m(W,g_0,h_\tau,\theta)
]
\bigg|_{\tau=0}
=
-\theta
\mathbb E[
Z_0\delta_2(X_{21})
].
\]
Since
\[
a_0(X_{21})
=
\mathbb E[Z_0\mid X_{21}],
\]
the second-step Riesz representer is
\[
\alpha_{02}(X_{21};\theta)
=
-\theta a_0(X_{21}).
\]

Next consider the first-step component $q_0$. For
$q_\tau=q_0+\tau\delta_q$, with $\delta_q\in\mathcal H$,
\begin{align*}
\frac{d}{d\tau}
\mathbb E[
m(W,(g_{01},q_\tau),h_0,\theta)
]
\bigg|_{\tau=0}
&=
-
\mathbb E\left[
\delta_q(X_{11},X_{12})
\{Y_2-\theta h_0(X_{21})\}
\right]
\\
&=
-
\mathbb E\left[
\delta_q(X_{11},X_{12})
\pi_0(X_{11},X_{12};\theta)
\right],
\end{align*}
where the last equality follows from the definition of
$\pi_0(\cdot;\theta)$ as the $L_2$ projection of
$Y_2-\theta h_0(X_{21})$ onto $\mathcal H$. Hence
\[
\alpha_{01q}(X_{11},X_{12};\theta)
=
-\pi_0(X_{11},X_{12};\theta).
\]

Finally, consider the first-step regression $g_{01}$. A perturbation
$g_{01,\tau}=g_{01}+\tau\delta_{11}$ induces
\[
D_{g_{01}}h_0[\delta_{11}]
=
-\delta_{11}(X_{11}),
\]
because $X_{11}$ is contained in $X_{21}$. Therefore, the total
derivative of the expected identifying moment with respect to $g_{01}$
is
\[
\theta
\mathbb E[
Z_0\delta_{11}(X_{11})
]
=
0,
\]
where the equality follows because
$L_2(X_{11})\subseteq\mathcal H$ and $Z_0\perp\mathcal H$.
Thus
\[
\alpha_{011}(X_{11};\theta)=0.
\]

Substituting these Riesz representers into the locally robust moment
of Theorem~\ref{thm:lr} gives
\begin{align*}
\psi(W,g_0,h_0,\alpha_0,\theta)
&=
Z_0
\{Y_2-\theta h_0(X_{21})\}
\\
&\quad
-\theta a_0(X_{21})
\{Y_1-g_{01}(X_{11})-h_0(X_{21})\}
\\
&\quad
-\pi_0(X_{11},X_{12};\theta)
\{R-q_0(X_{11},X_{12})\},
\end{align*}
which is \eqref{eq:prod_lr}.
$\square$.

\bigskip \noindent
\textbf{Proof of Proposition \ref{prop:iop}:}
Let $\delta_1\in L_2(A,C)$ be a first-step direction, $g_\tau=g_0+\tau\delta_1$, and
\[
\bar\delta_{\mathcal{R}}(C)=\sum_{b\in\mathcal{R}}p_{\mathcal{R}}(b)\,\delta_1(b,C)
\]
the direction averaged over the reference ages. Since $\tilde Y(g_\tau)=\tilde Y(g_0)+\tau\{\bar\delta_{\mathcal{R}}(C)-\delta_1(A,C)\}$, the derivative of corrected income is
\[
D_gS(g_0)[\delta_1]=\dot S\,\big\{\bar\delta_{\mathcal{R}}(C)-\delta_1(A,C)\big\},
\]
with $\dot S=1$ if $S(g)=\tilde Y(g)$ and $\dot S=e^{\tilde Y}$ if $S(g)=e^{\tilde Y(g)}$. The term $\bar\delta_{\mathcal{R}}(C)$ evaluates the direction at the reference ages, and the weights $w_{\mathcal{R}}$ express it at the observed ages. Under overlap, $w_{\mathcal{R}}\leq\max_{b\in\mathcal{R}}p_{\mathcal{R}}(b)/\underline{\pi}$ is bounded, and for every $c$,
\[
\mathbb{E}\big[w_{\mathcal{R}}(A,C)\,\delta_1(A,C)\mid C=c\big]
=
\sum_{a}p_0(a\mid c)\,\frac{\mathbf{1}\{a\in\mathcal{R}\}\,p_{\mathcal{R}}(a)}{p_0(a\mid c)}\,\delta_1(a,c)
=
\bar\delta_{\mathcal{R}}(c).
\]
Hence, by iterated expectations, for every $f\in L_2(C)$,
\begin{equation}\label{eq:iop_weights}
\mathbb{E}\big[f(C)\,\bar\delta_{\mathcal{R}}(C)\big]
=
\mathbb{E}\big[f(C)\,w_{\mathcal{R}}(A,C)\,\delta_1(A,C)\big],
\end{equation}
and, taking $\delta_1\equiv1$, $\mathbb{E}[w_{\mathcal{R}}(A,C)\mid C]=1$, since the reference weights sum to one.

By Theorem~\ref{thm:lr}, it suffices to find, for each component of $m$, the representers $\alpha_{02}$ and $\alpha_{01}$ satisfying \eqref{eq:riesz2} and \eqref{eq:riesz1}. We first consider the second-step nuisance. Only the numerator components of $m$ depend on $h$, and for every $\delta_2\in L_2(C)$ their derivative is $\mathbb{E}[\kappa'(h_0(C))\,\delta_2(C)]$. Hence
\[
\alpha_{02}(C)=\big(\kappa'(h_0(C))^{\top},\,0^{\top},\,0\big)^{\top},
\]
whose components belong to $L_2(C)$ by the moment conditions.

Next consider the first step in the numerator components, which depend on $g$ only through $h$. Since $\Delta_2=L_2(C)$, Remark~\ref{rem:linear} gives $\dot h_g(\delta_1)(C)=\mathbb{E}[D_gS(g_0)[\delta_1]\mid C]$, so the total first-step derivative of these components is
\begin{align*}
\dot m_{g,\mathrm{tot}}(\delta_1)
&=\mathbb{E}\Big[\kappa'\big(h_0(C)\big)\,\mathbb{E}\big[D_gS(g_0)[\delta_1]\mid C\big]\Big]\\
&=\mathbb{E}\Big[\kappa'\big(h_0(C)\big)\,\dot S\,\big\{\bar\delta_{\mathcal{R}}(C)-\delta_1(A,C)\big\}\Big]\\
&=\mathbb{E}\Big[Q_N(A,C)\,\big\{\bar\delta_{\mathcal{R}}(C)-\delta_1(A,C)\big\}\Big],
\end{align*}
where the second equality uses iterated expectations, because $\kappa'(h_0(C))$ is a function of $C$, and the third conditions on $(A,C)$, because $\bar\delta_{\mathcal{R}}(C)-\delta_1(A,C)$ is a function of $(A,C)$. Since $\bar\delta_{\mathcal{R}}$ is a function of $C$, $\mathbb{E}[Q_N\,\bar\delta_{\mathcal{R}}(C)]=\mathbb{E}\big[\mathbb{E}[Q_N\mid C]\,\bar\delta_{\mathcal{R}}(C)\big]$, and \eqref{eq:iop_weights} with $f=\mathbb{E}[Q_N\mid C]$ yields
\[
\dot m_{g,\mathrm{tot}}(\delta_1)
=
\mathbb{E}\Big[\big\{w_{\mathcal{R}}(A,C)\,\mathbb{E}[Q_N\mid C]-Q_N(A,C)\big\}\,\delta_1(A,C)\Big].
\]

The denominator components do not depend on $h$, so their total first-step derivative is the direct effect. By the same argument,
\begin{align*}
\dot m_{g,\mathrm{tot}}(\delta_1)
&=\mathbb{E}\Big[\kappa'\big(S(g_0)\big)\,\dot S\,\big\{\bar\delta_{\mathcal{R}}(C)-\delta_1(A,C)\big\}\Big]\\
&=\mathbb{E}\Big[\big\{w_{\mathcal{R}}(A,C)\,\mathbb{E}[Q_D\mid C]-Q_D(A,C)\big\}\,\delta_1(A,C)\Big].
\end{align*}
Finally, the last component of $m$ depends neither on $g$ nor on $h$, so both of its representers are zero. Since $w_{\mathcal{R}}$ is bounded and $Q_N$, $Q_D$, $\mathbb{E}[Q_N\mid C]$ and $\mathbb{E}[Q_D\mid C]$ have finite second moments, by the moment conditions and conditional Jensen's inequality, the components of $\alpha_{01}$ belong to $L_2(A,C)$. Theorem~\ref{thm:lr} then implies that $\psi$ is the locally robust moment for $(\bar\kappa_0,\theta_0)$.
$\square$

\bigskip \noindent
\textbf{Proof of Corollary \ref{cor:iop}:}
In both cases, write $m=(m_1,m_2)'$, where $m_1$ identifies $\theta_0$ and $m_2$ identifies $\mu_0$, and apply Theorem~\ref{thm:lr} to each component. We first consider the variance of logs, for which
\[
m_1=\big(h(C)-\mu\big)^2-\theta\big(\tilde Y(g)-\mu\big)^2,
\qquad
m_2=\tilde Y(g)-\mu,
\]
$S(g)=\tilde Y(g)$ and $\dot S=1$, so that $\dot h_g(\delta_1)(C)=\mathbb{E}[\bar\delta_{\mathcal{R}}(C)-\delta_1(A,C)\mid C]$. The derivative of $m_1$ with respect to $h$ is $2(h_0(C)-\mu)$, which yields the second-step correction $2(h(C)-\mu)(\tilde Y(g)-h(C))$. The total first-step derivative of $m_1$ is
\begin{align*}
\dot m_{g,\mathrm{tot}}(\delta_1)
&=2\,\mathbb{E}\big[(h_0(C)-\mu)\,\dot h_g(\delta_1)(C)\big]\\
&\quad-2\theta\,\mathbb{E}\big[(\tilde Y-\mu)\big\{\bar\delta_{\mathcal{R}}(C)-\delta_1(A,C)\big\}\big].
\end{align*}
By iterated expectations, the first term equals $2\,\mathbb{E}[(h_0(C)-\mu)\{\bar\delta_{\mathcal{R}}(C)-\delta_1(A,C)\}]$. In the second, $\tilde Y-\mu=h_0(C)-\mu+\varepsilon$ with $\mathbb{E}[\varepsilon\mid A,C]=0$, and $\bar\delta_{\mathcal{R}}(C)-\delta_1(A,C)$ is a function of $(A,C)$, so the second term equals $\theta$ times the first. Applying \eqref{eq:iop_weights} with $f=h_0-\mu$ then gives
\[
\dot m_{g,\mathrm{tot}}(\delta_1)
=
\mathbb{E}\Big[2(1-\theta)\big(h_0(C)-\mu\big)\big(w_{\mathcal{R}}(A,C)-1\big)\,\delta_1(A,C)\Big],
\]
which identifies $\alpha_{01,1}$. The component $m_2$ does not depend on $h$, and by \eqref{eq:iop_weights} with $f=1$ its first-step derivative is
\[
\mathbb{E}\big[\bar\delta_{\mathcal{R}}(C)-\delta_1(A,C)\big]
=
\mathbb{E}\big[\{w_{\mathcal{R}}(A,C)-1\}\,\delta_1(A,C)\big],
\]
so that $\alpha_{01,2}=w_{\mathcal{R}}-1$.

For the mean log deviation,
\[
m_1=\ln\mu-\ln h(C)-\theta\big(\ln\mu-\tilde Y(g)\big),
\qquad
m_2=e^{\tilde Y(g)}-\mu,
\]
$S(g)=e^{\tilde Y(g)}$ and $\dot S=e^{\tilde Y}$. Since $\mathbb{E}[e^{\tilde Y}\bar\delta_{\mathcal{R}}(C)\mid C]=h_0(C)\,\bar\delta_{\mathcal{R}}(C)$ and $\mathbb{E}[e^{\tilde Y}\delta_1(A,C)\mid C]=\mathbb{E}[H_0(A,C)\,\delta_1(A,C)\mid C]$,
\[
\dot h_g(\delta_1)(C)=h_0(C)\,\bar\delta_{\mathcal{R}}(C)-\mathbb{E}\big[H_0(A,C)\,\delta_1(A,C)\mid C\big].
\]
The derivative of $m_1$ with respect to $h$ is $-1/h_0(C)$, which is bounded since $h_0(C)\geq c>0$, and yields the second-step correction $-\{e^{\tilde Y(g)}-h(C)\}/h(C)$. The total first-step derivative of $m_1$ combines the indirect effect through $h$ with the direct effect through $\theta\,\tilde Y(g)$:
\begin{align*}
\dot m_{g,\mathrm{tot}}(\delta_1)
&=-\mathbb{E}\bigg[\frac{\dot h_g(\delta_1)(C)}{h_0(C)}\bigg]+\theta\,\mathbb{E}\big[\bar\delta_{\mathcal{R}}(C)-\delta_1(A,C)\big]\\
&=-\mathbb{E}\big[\bar\delta_{\mathcal{R}}(C)\big]+\mathbb{E}\bigg[\frac{H_0(A,C)}{h_0(C)}\,\delta_1(A,C)\bigg]+\theta\,\mathbb{E}\big[\bar\delta_{\mathcal{R}}(C)-\delta_1(A,C)\big]\\
&=\mathbb{E}\bigg[\Big\{\frac{H_0(A,C)}{h_0(C)}-w_{\mathcal{R}}(A,C)+\theta\big(w_{\mathcal{R}}(A,C)-1\big)\Big\}\,\delta_1(A,C)\bigg],
\end{align*}
where the second equality uses iterated expectations and the third applies \eqref{eq:iop_weights} with $f=1$. This identifies $\alpha_{01,1}$. The component $m_2$ does not depend on $h$, and its first-step derivative is
\begin{align*}
\dot m_{g,\mathrm{tot}}(\delta_1)
&=\mathbb{E}\big[e^{\tilde Y}\big\{\bar\delta_{\mathcal{R}}(C)-\delta_1(A,C)\big\}\big]\\
&=\mathbb{E}\big[h_0(C)\,\bar\delta_{\mathcal{R}}(C)\big]-\mathbb{E}\big[H_0(A,C)\,\delta_1(A,C)\big]\\
&=\mathbb{E}\Big[\big\{w_{\mathcal{R}}(A,C)\,h_0(C)-H_0(A,C)\big\}\,\delta_1(A,C)\Big],
\end{align*}
where the last equality applies \eqref{eq:iop_weights} with $f=h_0$. This identifies $\alpha_{01,2}$. Assumption~\ref{ass:lr} holds for the variance of logs because, by Jensen's inequality and overlap, $\|\bar\delta_{\mathcal{R}}\|_2^2\leq\underline{\pi}^{-1}\max_{b\in\mathcal{R}}p_{\mathcal{R}}(b)\,\|\delta_1\|_2^2$, so $\tilde Y(g)$ is affine and continuous in $g$ and the moment is quadratic. For the mean log deviation, finite support gives $\|\delta_1\|_\infty\leq p_{\min}^{-1/2}\|\delta_1\|_2$, with $p_{\min}$ the smallest probability of a support point, so the bound $|e^x-1-x|\leq x^2e^{|x|}/2$ makes $g\mapsto e^{\tilde Y(g)}$ Fr\'echet differentiable into $L_2$, and $h_0\geq c$ does the same for $h\mapsto\ln h(C)$. In both cases, Theorem~\ref{thm:lr} gives the stated locally robust moments.
$\square$

\bigskip \noindent
\textbf{Proof of Lemma \ref{lem:consistency}:} Fix a fold $\ell$ and let $I_\ell^c$ denote the complement of $\mathcal{I}_\ell$. Since $\hat{g}_\ell$, $\hat{h}_\ell$, and $\hat{\alpha}_\ell$ depend only on $I_\ell^c$, observations $\{W_i : i \in \mathcal{I}_\ell\}$ are i.i.d.\ conditionally on $I_\ell^c$. Decompose
\[\hat{\psi}_{i\ell}(\theta_0) - \psi(W_i, g_0, h_0, \alpha_0, \theta_0) = \hat{R}_{1i\ell} + \hat{R}_{2i\ell} + \hat{\Delta}_{i\ell},\]
where
\begin{align*}
\hat{R}_{1i\ell} &\equiv \psi(W_i, \hat{g}_\ell, \hat{h}_\ell, \alpha_0, \theta_0) - \psi(W_i, g_0, h_0, \alpha_0, \theta_0),\\
\hat{R}_{2i\ell} &\equiv \psi(W_i, g_0, h_0, \hat{\alpha}_\ell, \theta_0) - \psi(W_i, g_0, h_0, \alpha_0, \theta_0),\\
\hat{\Delta}_{i\ell} &\equiv \hat{\psi}_{i\ell}(\theta_0) - \psi(W_i, \hat{g}_\ell, \hat{h}_\ell, \alpha_0, \theta_0) - \psi(W_i, g_0, h_0, \hat{\alpha}_\ell, \theta_0) + \psi(W_i, g_0, h_0, \alpha_0, \theta_0).
\end{align*}
It suffices to show that $n^{-1}\sum_{i \in \mathcal{I}_\ell}(\hat{R}_{1i\ell} + \hat{R}_{2i\ell} + \hat{\Delta}_{i\ell}) \overset{p}{\to} 0$ for each $\ell$. To control $\hat R_{1i\ell}$, define
\[
\hat R_{1\ell}(w)
=
\psi(w,\hat g_\ell,\hat h_\ell,\alpha_0,\theta_0)
-
\psi(w,g_0,h_0,\alpha_0,\theta_0),
\]
so that $\hat R_{1i\ell}=\hat R_{1\ell}(W_i)$.
For a vector-valued function $f$, write
\[
\|f\|_r
=
\left(\int \|f(w)\|^r\,dF_0(w)\right)^{1/r},
\qquad r=1,2.
\]
At fixed $(\alpha_0,\theta_0)$, the correction
$\phi=\phi_1+\phi_2$ is additively separable in $g$ and $h$:
\begin{align*}
&\phi(w,\hat g_\ell,\hat h_\ell,\alpha_0,\theta_0)
-
\phi(w,g_0,h_0,\alpha_0,\theta_0)
\\
&\quad =
\phi(w,\hat g_\ell,h_0,\alpha_0,\theta_0)
-
\phi(w,g_0,h_0,\alpha_0,\theta_0)
\\
&\qquad\quad +
\phi(w,g_0,\hat h_\ell,\alpha_0,\theta_0)
-
\phi(w,g_0,h_0,\alpha_0,\theta_0).
\end{align*}
Consequently, the triangle inequality and
$\|f\|_1\leq\|f\|_2$ give
\begin{align*}
\|\hat R_{1\ell}\|_1
&\leq
\bigl\|
m(\cdot,\hat g_\ell,\hat h_\ell,\theta_0)
-
m(\cdot,g_0,h_0,\theta_0)
\bigr\|_1
\\
&\quad+
\bigl\|
\phi(\cdot,\hat g_\ell,h_0,\alpha_0,\theta_0)
-
\phi(\cdot,g_0,h_0,\alpha_0,\theta_0)
\bigr\|_2
\\
&\quad+
\bigl\|
\phi(\cdot,g_0,\hat h_\ell,\alpha_0,\theta_0)
-
\phi(\cdot,g_0,h_0,\alpha_0,\theta_0)
\bigr\|_2
=
o_p(1),
\end{align*}
where the first term converges to zero by
Assumption~\ref{ass:consistency}$(ii)$ and the remaining
terms converge to zero by
Assumption~\ref{ass:consistency}$(iv)$.

Let $\mathcal F_{-\ell}$ denote the sigma-field generated by
the training observations outside $\mathcal I_\ell$, and let
$n_\ell=|\mathcal I_\ell|$.
Conditionally on $\mathcal F_{-\ell}$,
$\hat R_{1\ell}$ is fixed and each held-out observation has
distribution $F_0$. Hence
\[
\mathbb E\left[
\frac{1}{n}\sum_{i\in\mathcal I_\ell}
\|\hat R_{1\ell}(W_i)\|
\,\middle|\,\mathcal F_{-\ell}
\right]
=
\frac{n_\ell}{n}\|\hat R_{1\ell}\|_1.
\]
By the triangle inequality and conditional Markov inequality,
for every $\varepsilon>0$,
\[
\Pr\left(
\left\|
\frac{1}{n}\sum_{i\in\mathcal I_\ell}
\hat R_{1i\ell}
\right\|>\varepsilon
\right)
\leq
\mathbb E\left[
\min\left\{
1,\frac{n_\ell}{n\varepsilon}\|\hat R_{1\ell}\|_1
\right\}
\right]
\longrightarrow 0.
\]
The last convergence follows because the integrand is bounded
by one and converges to zero in probability.
Therefore,
\[
\frac{1}{n}\sum_{i\in\mathcal I_\ell}
\hat R_{1i\ell}
=
o_p(1).
\]
\par
Turning attention to $\hat{R}_{2i\ell}$, by the zero-mean property of the influence functions, $\mathbb{E}[\hat{R}_{2i\ell} \mid I_\ell^c] = \mathbb{E}[\phi(W, g_0, h_0, \hat{\alpha}_\ell, \theta_0)] = 0$. Condition~$(iv)$ gives $\mathbb{E}[\hat{R}_{2i\ell}^2 \mid I_\ell^c] \overset{p}{\to} 0$, so $n^{-1}\sum_{i \in \mathcal{I}_\ell}\hat{R}_{2i\ell} \overset{p}{\to} 0$ by the conditional Markov inequality. Since $n^{-1}\sum_{i \in \mathcal{I}_\ell}\hat{\Delta}_{i\ell} = o_p(1)$ by condition~$(v)$, by the triangle inequality,
\[\frac{1}{n}\sum_{i \in \mathcal{I}_\ell}\hat{\psi}_{i\ell}(\theta_0) \overset{p}{\to} \mathbb{E}[\psi(W, g_0, h_0, \alpha_0, \theta_0)] = \bar{m}(\gamma_0, \theta_0) = 0,\]
where the first equality follows by the Law of Large Numbers and the second by the identifying condition. Summing across folds gives $\hat{\psi}(\theta_0) \overset{p}{\to} 0$. By Assumption~\ref{ass:consistency}$(iii)$ and conditional
Markov's inequality applied fold by fold,
\[
\sup_{\theta\in\Theta}
\left\|
\hat{\psi}(\theta)
-\frac{1}{n}\sum_{i=1}^n
\psi(W_i,g_0,h_0,\alpha_0,\theta)
\right\|=o_p(1).
\]
Compactness of $\Theta$, integrability at $\theta_0$, and the
continuity bound in Assumption~\ref{ass:consistency}$(iii)$
give a uniform law of large numbers for the true-nuisance
score. Since
$\mathbb{E}[\psi(W,g_0,h_0,\alpha_0,\theta)]
=\bar{m}(\gamma_0,\theta)$ by the projection conditions,
\[
\sup_{\theta\in\Theta}
\|\hat{\psi}(\theta)-\bar{m}(\gamma_0,\theta)\|
\overset{p}{\to}0.
\]
Together with Assumption~\ref{ass:consistency}$(vi)$, this
implies uniform convergence of the GMM criterion to
$\bar{m}(\gamma_0,\theta)'\Upsilon\bar{m}(\gamma_0,\theta)$.
This limit is continuous and uniquely minimized at $\theta_0$
by Assumption~\ref{ass:consistency}$(i)$.
The usual argmin argument on compact $\Theta$ therefore gives
$\hat{\theta}\overset{p}{\to}\theta_0$. $\square$

\bigskip \noindent
\textbf{Proof of Theorem \ref{thm:normality}:} The challenge lies in establishing the asymptotic equivalence
\begin{align}
\label{eq:asy_equiv}
\sqrt{n}\hat{\psi}(\theta_0) = \frac{1}{\sqrt{n}}\sum_{i=1}^n \psi(W_i, g_0, h_0, \alpha_0, \theta_0) + o_p(1),
\end{align}
which requires showing that $n^{-1/2}\sum_{i \in \mathcal{I}_\ell}(\hat{R}_{1i\ell} + \hat{R}_{2i\ell} + \hat{\Delta}_{i\ell}) \overset{p}{\to} 0$ for each $\ell$, using the same decomposition as in Lemma~\ref{lem:consistency}. The terms $\hat{R}_{2i\ell}$ and $\hat{\Delta}_{i\ell}$ carry over directly from Lemma~\ref{lem:consistency}: the zero-mean property of the influence functions and Assumption~\ref{ass:consistency}$(iv)$ give $n^{-1/2}\sum_{i \in \mathcal{I}_\ell}\hat{R}_{2i\ell} \overset{p}{\to}0$ by the conditional variance bound and Chebyshev's inequality, since $n_\ell=O(n)$, and $n^{-1/2}\sum_{i \in \mathcal{I}_\ell}\hat{\Delta}_{i\ell} = o_p(1)$ by Assumption~\ref{ass:consistency}$(v)$. It remains to show that $n^{-1/2}\sum_{i \in \mathcal{I}_\ell}\hat{R}_{1i\ell} \overset{p}{\to} 0$, which requires additional argument at the $\sqrt{n}$ rate.\par
Since the population moment vanishes at $\theta_0$,
\[
\mathbb{E}[\hat{R}_{1i\ell}\mid I_\ell^c]
=
\bar{\psi}(\hat{g}_\ell,\hat{h}_\ell,\alpha_0,\theta_0)
=
o_p(n^{-1/2})
\]
by Assumption~\ref{ass:normality}$(vi)$.
Moreover, Assumptions~\ref{ass:consistency}$(iv)$ and
\ref{ass:normality}$(v)$, together with the additive decomposition
of $\phi$ used in the proof of Lemma~\ref{lem:consistency}, imply
\[
\mathbb{E}[\|\hat{R}_{1i\ell}\|^2\mid I_\ell^c]
=
\int\|\hat{R}_{1\ell}(w)\|^2\,dF_0(w)
=
o_p(1).
\]
Conditional independence within $\mathcal{I}_\ell$ and
Chebyshev's inequality therefore give, with
$n_\ell=|\mathcal{I}_\ell|=O(n)$,
\[
\frac{1}{\sqrt{n}}\sum_{i\in\mathcal{I}_\ell}\hat{R}_{1i\ell}
=
\frac{n_\ell}{\sqrt{n}}\,
\mathbb{E}[\hat{R}_{1i\ell}\mid I_\ell^c]
+
o_p(1)
=
o_p(1).
\]
Together with the bounds for $\hat{R}_{2i\ell}$ and
$\hat{\Delta}_{i\ell}$, this establishes
equation~\eqref{eq:asy_equiv}.\par
By the multivariate central limit theorem, which does not require $\Psi$ to be nonsingular,
\[\frac{1}{\sqrt{n}}\sum_{i=1}^n \psi(W_i, g_0, h_0, \alpha_0, \theta_0) \xrightarrow{d} \mathcal{N}(0, \Psi),\]
where $\Psi = \mathbb{E}[\psi(W, g_0, h_0, \alpha_0, \theta_0)\psi(W, g_0, h_0, \alpha_0, \theta_0)']$. To derive the asymptotic distribution of $\sqrt{n}(\hat{\theta} - \theta_0)$, let $\hat{G}(\theta) \equiv \partial\hat{\psi}(\theta)/\partial\theta$. By consistency and Assumption~\ref{ass:normality}$(ii)$, $\hat{\theta}\in\operatorname{int}(\Theta)$ with probability approaching one. On this event, the GMM first-order condition gives $\hat{G}(\hat{\theta})'\hat{\Upsilon}\hat{\psi}(\hat{\theta})=0$. Expanding $\hat{\psi}(\hat{\theta})$ around $\theta_0$ by the mean value theorem yields $0 = \hat{G}(\hat{\theta})'\hat{\Upsilon}\hat{\psi}(\theta_0) + \hat{G}(\hat{\theta})'\hat{\Upsilon}\hat{G}(\bar{\theta})(\hat{\theta} - \theta_0)$ for $\bar{\theta}$ between $\hat{\theta}$ and $\theta_0$, so that
\[\sqrt{n}(\hat{\theta} - \theta_0) = -\bigl(\hat{G}(\hat{\theta})'\hat{\Upsilon}\hat{G}(\bar{\theta})\bigr)^{-1}\hat{G}(\hat{\theta})'\hat{\Upsilon}\sqrt{n}\hat{\psi}(\theta_0).\] Here and below, $\hat G(\bar\theta)$ is understood as shorthand for the
integrated Jacobian along the line segment joining $\theta_0$ and
$\hat\theta$.
By consistency of $\hat{\theta}$ (Lemma~\ref{lem:consistency}) and Assumptions~\ref{ass:normality}$(iii)$--$(iv)$, $\hat{G}(\hat{\theta}) \overset{p}{\to} G$ and $\hat{G}(\bar{\theta}) \overset{p}{\to} G$, where $G=\mathbb{E}[ \partial_\theta \psi(W,g_0,h_0,\alpha_0,\theta_0)]$. Together with $\hat{\Upsilon} \overset{p}{\to} \Upsilon$ and equation~(\ref{eq:asy_equiv}), Slutsky's theorem gives $\sqrt{n}(\hat{\theta} - \theta_0) \xrightarrow{d} \mathcal{N}(0, V)$, where $V = (G'\Upsilon G)^{-1}(G'\Upsilon\Psi\Upsilon G)(G'\Upsilon G)^{-1}$ and $G'\Upsilon G$ is nonsingular by Assumption~\ref{ass:normality}$(ii)$. For variance estimation, let
$\psi_i=\psi(W_i,g_0,h_0,\alpha_0,\theta_0)$,
$\bar{\Psi}=n^{-1}\sum_{i=1}^n\psi_i\psi_i'$, and
\[
\hat{\Psi}
=
\frac{1}{n}\sum_{\ell=1}^L\sum_{i\in\mathcal{I}_\ell}
\hat{\psi}_{i\ell}(\hat{\theta})
\hat{\psi}_{i\ell}(\hat{\theta})'.
\]
The preceding mean-square bounds and conditional Markov's
inequality give
\[
\frac{1}{n}\sum_{\ell=1}^L\sum_{i\in\mathcal{I}_\ell}
\|\hat{\psi}_{i\ell}(\theta_0)-\psi_i\|^2=o_p(1).
\]
To replace $\theta_0$ by $\hat{\theta}$, set
$D_{i\ell}=\|\partial_\theta\hat{\psi}_{i\ell}(\theta_0)\|
+d(W_i)$.
By Assumption~\ref{ass:normality}$(iii)$ and consistency,
$D_{i\ell}$ bounds the Jacobian norm along the segment joining
$\theta_0$ and $\hat{\theta}$ with probability approaching one.
The $L_1$ approximation in
Assumption~\ref{ass:normality}$(iv)$, conditional Markov's
inequality, and integrability of the true Jacobian and $d(W)$
give
\[
\frac{1}{n}\sum_{\ell=1}^L\sum_{i\in\mathcal{I}_\ell}
D_{i\ell}=O_p(1),
\qquad
\frac{1}{n}\max_{\ell,\,i\in\mathcal{I}_\ell}D_{i\ell}=o_p(1).
\]
For the maximum, use that the maximum of an i.i.d.\ integrable
sequence is $o_p(n)$ and that the sum of the absolute
Jacobian-approximation errors is $o_p(n)$.
Since $\sqrt{n}(\hat{\theta}-\theta_0)=O_p(1)$,
\begin{align*}
&\frac{1}{n}\sum_{\ell=1}^L\sum_{i\in\mathcal{I}_\ell}
\|\hat{\psi}_{i\ell}(\hat{\theta})
-\hat{\psi}_{i\ell}(\theta_0)\|^2\\
&\quad\leq
n\|\hat{\theta}-\theta_0\|^2
\left(\frac{\max_{\ell,\,i\in\mathcal{I}_\ell}D_{i\ell}}{n}\right)
\left(\frac{1}{n}\sum_{\ell=1}^L
\sum_{i\in\mathcal{I}_\ell}D_{i\ell}\right)
=o_p(1).
\end{align*}
Consequently,
\[
A_n\coloneqq
\frac{1}{n}\sum_{\ell=1}^L\sum_{i\in\mathcal{I}_\ell}
\|\hat{\psi}_{i\ell}(\hat{\theta})-\psi_i\|^2=o_p(1).
\]
Expanding the outer products and applying Cauchy-Schwarz yields
\[
\|\hat{\Psi}-\bar{\Psi}\|
\leq
2A_n^{1/2}
\left(\frac{1}{n}\sum_{i=1}^n\|\psi_i\|^2\right)^{1/2}
+A_n
=o_p(1).
\]
By Assumption~\ref{ass:normality}$(i)$ and the law of large
numbers, $\bar{\Psi}\overset{p}{\to}\Psi$, hence
$\hat{\Psi}\overset{p}{\to}\Psi$.
Together with $\hat{G}(\hat{\theta})\overset{p}{\to}G$
and $\hat{\Upsilon}\overset{p}{\to}\Upsilon$, continuity
of the sandwich formula gives
$\hat{V}\overset{p}{\to}V$, and Slutsky's theorem gives the studentized limit whenever $c'Vc>0$. $\square$

\bigskip \noindent
\textbf{Proof of Corollary \ref{cor:plm_normality}:} It suffices to verify Assumptions~\ref{ass:consistency} and \ref{ass:normality} and invoke Theorem~\ref{thm:normality}.

\medskip \noindent\textit{Assumption \ref{ass:consistency}(i).}
For applying Theorem~\ref{thm:normality}, take $\Theta$ to be
a compact neighborhood of $\theta_0$.
Cross-fitting, H\"older's inequality, and
Assumption~\ref{ass:plm}$(ii)$--$(iii)$ give
$\hat{B}\overset{p}{\to}B$ and $\hat{C}\overset{p}{\to}C$.
Since $B$ is nonsingular,
$\hat{B}^{-1}\hat{C}\overset{p}{\to}\theta_0$.
Thus the explicit estimator lies in $\operatorname{int}(\Theta)$
and minimizes the GMM criterion with probability approaching one.
The population moment is $C-B\theta$, whose unique zero is
$\theta_0=B^{-1}C$.

\medskip \noindent\textit{Assumption \ref{ass:consistency}(ii).}
Let $\nu=X_3-h_{0X}(X_2)$ and
$U=g_0(X_1)-h_{0Y}(X_2)-\theta_0'\nu$.
Conditional Jensen's inequality and
Assumption~\ref{ass:plm}$(ii)$ imply that
$g_0,h_0,\alpha_{01},U,\nu$, and
$\epsilon(W,g_0)=Y-g_0(X_1)$ have finite fourth moments.
At the true nuisances,
$\psi=U\nu+\alpha_{01}(X_1)\epsilon(W,g_0)$, so
H\"older's inequality gives
\[
\|\psi\|_2
\leq
\|U\|_4\|\nu\|_4
+\|\alpha_{01}\|_4\|\epsilon(W,g_0)\|_4
<\infty.
\]
For each fixed $\theta\in\Theta$, expanding the PLM moment
and applying H\"older's inequality gives
\begin{align*}
&\|m(\cdot,\hat{g}_\ell,\hat{h}_\ell,\theta)
-m(\cdot,g_0,h_0,\theta)\|_2\\
&\quad\leq
C(1+\|\hat{h}_\ell-h_0\|_4)
\bigl(\|\hat{g}_\ell-g_0\|_4
+\|\hat{h}_\ell-h_0\|_4\bigr)
=o_p(1)
\end{align*}
by Assumption~\ref{ass:plm}$(iii)$.
The required $L_1$ conclusions follow from
$\|f\|_1\leq\|f\|_2$.

\medskip \noindent\textit{Assumption \ref{ass:consistency}(iii).}
The true score is Lipschitz in $\theta$ with
$d(W)=\|\nu\|^2$, which is integrable.
Let $\hat{\nu}_\ell=X_3-\hat{h}_{X\ell}(X_2)$.
Both scores are affine in $\theta$, with slopes
$-\hat{\nu}_\ell\hat{\nu}_\ell'$ and $-\nu\nu'$, respectively.
By Cauchy-Schwarz,
\[
\|\hat{\nu}_\ell\hat{\nu}_\ell'-\nu\nu'\|_1
\leq
(\|\hat{\nu}_\ell\|_2+\|\nu\|_2)
\|\hat{h}_{X\ell}-h_{0X}\|_2
=o_p(1).
\]
The full score difference at $\theta_0$ also converges to zero
in $L_1$ by H\"older's inequality and
Assumption~\ref{ass:plm}$(ii)$--$(iii)$.
Affine dependence on $\theta$ and compactness of $\Theta$
therefore yield the required uniform $L_1$ convergence.
\medskip \noindent\textit{Assumption \ref{ass:consistency}(iv).}
For the PLM, $\phi_2=0$ and $\hat{\alpha}_{2\ell}=0$.
By H\"older's inequality,
\begin{align*}
\|\alpha_{01}(\hat{g}_\ell-g_0)\|_2
&\leq
\|\alpha_{01}\|_4\|\hat{g}_\ell-g_0\|_4
=o_p(1),\\
\|(\hat{\alpha}_{01\ell}-\alpha_{01})\epsilon(W,g_0)\|_2
&\leq
\|\hat{\alpha}_{01\ell}-\alpha_{01}\|_4
\|\epsilon(W,g_0)\|_4
=o_p(1)
\end{align*}
by Assumption~\ref{ass:plm}$(ii)$--$(iii)$.
These bounds control the differences in $g$ and $\alpha$;
the difference in $h$ is identically zero.

\medskip \noindent\textit{Assumption \ref{ass:consistency}(v).}
For the PLM, set $\hat{\alpha}_{2\ell}=\alpha_{02}=0$.
The interaction remainder is
\[
\hat{\Delta}_\ell(w)
=
-\bigl(\hat{\alpha}_{01\ell}(x_1)-\alpha_{01}(x_1)\bigr)
\bigl(\hat{g}_\ell(x_1)-g_0(x_1)\bigr).
\]
By Cauchy-Schwarz and Assumption~\ref{ass:plm}$(iv)$,
with $n_\ell=|\mathcal{I}_\ell|$,
\[
\mathbb{E}\left[
\frac{1}{\sqrt{n}}\sum_{i\in\mathcal{I}_\ell}
\|\hat{\Delta}_\ell(W_i)\|
\,\middle|\,I_\ell^c
\right]
\leq
\frac{n_\ell}{\sqrt{n}}\,
\|\hat{\alpha}_{01\ell}-\alpha_{01}\|_2
\|\hat{g}_\ell-g_0\|_2
=o_p(1).
\]
Conditional Markov's inequality therefore gives
$n^{-1/2}\sum_{i\in\mathcal{I}_\ell}
\hat{\Delta}_\ell(W_i)=o_p(1)$.

\medskip \noindent\textit{Assumption \ref{ass:consistency}(vi).} Assumed directly.

\medskip \noindent\textit{Assumption \ref{ass:normality}(i).}
The required finite second moment of $\psi$ was established
in the verification of Assumption~\ref{ass:consistency}$(ii)$.

\medskip \noindent\textit{Assumption \ref{ass:normality}(ii).}
Interiority follows from the choice of $\Theta$ above.
The Jacobian is $G = \mathbb{E}[\partial_\theta m]
= -\mathbb{E}[(X_3-h_{0X}(X_2))(X_3-h_{0X}(X_2))']
= -B$, which exists and $G'\Upsilon G$ is nonsingular
by Assumption \ref{ass:plm}(i).

\medskip\noindent\textit{Assumption \ref{ass:normality}(iii).} Since $m$ is linear in $\theta$, the Jacobian $\partial m/\partial\theta = -(X_3 - h_{0X}(X_2))(X_3 - h_{0X}(X_2))'$ is constant in $\theta$, so the Lipschitz condition holds trivially with $d(W) = 0$.

\medskip \noindent\textit{Assumption \ref{ass:normality}(iv).} Since $\partial m/\partial\theta = -(X_3 - h_{0X}(X_2))(X_3 - h_{0X}(X_2))'$ depends on $h$ only through $h_{0X}(X_2)$, by Cauchy-Schwarz,
\[
\int \left\|\frac{\partial m}{\partial\theta}(w, \hat{g}_\ell, \hat{h}_\ell, \theta_0) - \frac{\partial m}{\partial\theta}(w, g_0, h_0, \theta_0)\right\|\, dF_0(w) \leq C\|\hat{h}_{X\ell} - h_{0X}\|_2 \overset{p}{\to} 0
\]
by Assumption \ref{ass:plm}(iii).

\medskip \noindent\textit{Assumption \ref{ass:normality}(v).}
Mean-square convergence of the identifying moment follows
from the bound established under
Assumption~\ref{ass:consistency}$(ii)$, evaluated at $\theta_0$.
For the interaction remainder derived under
Assumption~\ref{ass:consistency}$(v)$, H\"older's inequality gives
\[
\|\hat{\Delta}_\ell\|_2
\leq
\|\hat{\alpha}_{01\ell}-\alpha_{01}\|_4
\|\hat{g}_\ell-g_0\|_4
=o_p(1)
\]
by Assumption~\ref{ass:plm}$(iii)$.

\medskip \noindent\textit{Assumption \ref{ass:normality}(vi).}
Suppressing covariate arguments, expansion of the PLM score
and the projection identities give
\begin{align*}
&\bar{\psi}(\hat{g}_\ell,\hat{h}_\ell,\alpha_0,\theta_0)\\
&\quad=
\mathbb{E}\Big[
(\hat{h}_{X\ell}-h_{0X})
\big\{(\hat{h}_{Y\ell}-h_{0Y})-(\hat{g}_\ell-g_0)\big\}
\Big]\\
&\qquad-
\mathbb{E}\Big[
(\hat{h}_{X\ell}-h_{0X})
(\hat{h}_{X\ell}-h_{0X})'
\Big]\theta_0.
\end{align*}
By Cauchy-Schwarz and $2ab\leq a^2+b^2$,
\[
\|\bar{\psi}(\hat{g}_\ell,\hat{h}_\ell,\alpha_0,\theta_0)\|
\leq
C\bigl(\|\hat{g}_\ell-g_0\|_2^2
+\|\hat{h}_\ell-h_0\|_2^2\bigr).
\]
Together with Assumption~\ref{ass:plm}$(iv)$, this verifies
condition~$(a)$ of Assumption~\ref{ass:normality}$(vi)$.
\medskip
All conditions of Assumptions~\ref{ass:consistency} and \ref{ass:normality} are satisfied. Theorem~\ref{thm:normality} then gives $\sqrt{n}(\hat{\theta} - \theta_0) \xrightarrow{d} \mathcal{N}(0, V)$. Since the PLM moment is exactly identified and $G=-B$,
the GMM variance simplifies, for any positive-definite limiting
weighting matrix $\Upsilon$, to
\[
V=G^{-1}\Psi G^{-T}
=B^{-1}\Psi B^{-1}.
\] Consistent variance estimation $\hat{V} \overset{p}{\to} V$ follows from Assumption~\ref{ass:normality}(v) and the continuous mapping theorem. $\square$

\end{appendices}
\end{document}